# What it takes to solve the Origin(s) of Life:
## An integrated review of techniques.

OoLEN (Origin of Life Early-career Network)[1];

Silke Asche[2†], Carla Bautista[3,4,5,†], David Boulesteix[6,†], Alexandre Champagne-Ruel[7,†], Cole Mathis[8,†,*], Omer Markovitch[9,†], Zhen Peng[10,11,†], Alyssa Adams[12], Avinash Vicholous Dass[13], Arnaud Buch[6], Eloi Camprubi[14], Enrico Sandro Colizzi[15], Stephanie Colón-Santos, Hannah Dromiack[16], Valentina Erastova[17], Amanda Garcia[10], Ghjuvan Grimaud[18,19], Aaron Halpern[20], Stuart A Harrison[20], Seán F. Jordan[21], Tony Z Jia[22,9], Amit Kahana[2], Artemy Kolchinsky[23], Odin Moron-Garcia[24], Ryo Mizuuchi[25], Jingbo Nan[26], Yuliia Orlova[27], Ben K. D. Pearce[28], Klaus Paschek[29], Martina Preiner[30], Silvana Pinna[31], Eduardo Rodríguez-Román[32,33], Loraine Schwander[34], Siddhant Sharma[35,9], Harrison B Smith[22,9], Andrey Vieira[36], Joana C. Xavier[37,38].

[1] www.oolen.org

[2] School of Chemistry, University of Glasgow, Glasgow, UK

[3] Institut de Biologie Intégrative et des Systèmes (IBIS), Université Laval, Québec, QC, Canada

[4] Département de Biologie, Faculté des Sciences et de Génie, Université Laval, Québec, QC, Canada

[5] Regroupement québécois de recherche sur la fonction, la structure et l'ingénierie des protéines (PROTEO), Université Laval, Québec, QC, Canada

[6] Laboratoire Génie des Procédés et Matériaux, CentraleSupélec, Gif-sur-Yvette, France

[7] Université de Montréal, Montréal, Canada

[8] Beyond Center for Fundamental Concepts in Science, Arizona State University, Tempe Arizona USA

[9] Blue Marble Space Institute of Science, Seattle, WA, USA

[10] Department of Bacteriology, University of Wisconsin–Madison, Madison, Wisconsin, USA

[11] Department of Geoscience, University of Wisconsin–Madison, Madison, Wisconsin, USA

[12] Cross Labs, Kyoto, Japan

[13] Origins Institute, Department of Physics and Astronomy, McMaster University, Hamilton, Canada

[14] School of Integrated Biological and Chemical Sciences, University of Texas Rio Grande Valley, Edinburg TX, USA

[15] Sainsbury Laboratory, University of Cambridge, United Kingdom

[16] Department of Physics, Arizona State University Tempe, AZ, USA

[17] School of Chemistry, University of Edinburgh, Joseph Black Building, Edinburgh, United Kingdom

[18]APC Microbiome Ireland, University College Cork, Co. Cork, Ireland;

[19]Food Biosciences Department, Teagasc Food Research Centre, Moorepark, Fermoy, Co. Cork, Ireland

[20]Department of Genetics, Evolution and Environment, University College London, London, UK

[21]Department of Life Sciences, Atlantic Technological University, ATU Sligo, Sligo, Ireland

[22]Earth-Life Science Institute, Tokyo Institute of Technology, Meguro-ku, Tokyo, Japan

[23]ICREA-Complex Systems Lab, Universitat Pompeu Fabra, Barcelona, Spain

[24]Functional and Evolutionary Ecology Department, Estación Experimental de Zonas Áridas (EEZA-CSIC), Almería, Spain

[25]Department of Electrical Engineering and Bioscience, Faculty of Science and Engineering, Waseda University, Shinjuku, Tokyo, Japan

[26]Southern University of Science and Technology, Department of Ocean Science and Engineering, Shenzhen, China

[27]University of Amsterdam, Swammerdam Institute of Life Sciences, Amsterdam, The Netherlands

[28]Johns Hopkins University, Baltimore, MD, USA

[29]Max Planck Institute for Astronomy, Heidelberg, Germany

[30]Microcosm Earth Center, Max Planck Institute for Terrestrial Microbiology and Philipps-University Marburg, Marburg, Germany

[31]Institut de science et d'ingénierie supramoléculaires (ISIS), Université de Strasbourg, Strasbourg, France

[32]Department of Biology, Emory University, Atlanta, GA, USA

[33]Center for Microbiology and Cell Biology, IVIC, Caracas, Venezuela

[34]Institute of Molecular Evolution, Biology Department, Math.-Nat. Faculty, Heinrich-Heine-Universität, Düsseldorf, Germany

[35]School of Chemistry, University of New South Wales, Sydney, Australia

[36]Linnaeuskade 47 2, 1098BK, Amsterdam, The Netherlands

[37]Dayhoff Labs, London, UK

[38]Department of Chemistry, Imperial College London, London, UK

[†]These authors contributed equally.

[.]Please address correspondence to cole.mathis@asu.edu

# Abstract

Understanding the origin(s) of life (OoL) is a fundamental challenge for science in the 21st century. Research on OoL spans many disciplines, including chemistry, physics, biology, planetary sciences, computer science, mathematics and philosophy. The sheer number of different scientific perspectives relevant to the problem has resulted in the coexistence of diverse tools, techniques, data, and software in OoL studies. This has made communication between the disciplines relevant to the OoL extremely difficult because the interpretation of data, analyses, or standards of evidence can vary dramatically. Here, we hope to bridge this wide field of study by providing common ground via the consolidation of tools and techniques rather than positing a unifying view on how life emerges. We review the common tools and techniques that have been used significantly in OoL studies in recent years. In particular, we aim to identify which information is most relevant for comparing and integrating the results of experimental analyses into mathematical and computational models. This review aims to provide a baseline expectation and understanding of technical aspects of origins research, rather than being a primer on any particular topic. As such, it spans broadly — from analytical chemistry to mathematical models — and highlights areas of future work that will benefit from a multidisciplinary approach to tackling the mystery of life's origin. Ultimately, we hope to empower a new generation of OoL scientists by reviewing how they can investigate life's origin, rather than dictating how to think about the problem.

# Contents

# 1.Introduction

The question of how life began on Earth is one of the oldest posed by humankind. For millennia, the seemingly ethereal nature of living beings was attributed to supernatural forces that powered inanimate matter with unearthly properties, making it then living. Much of these concepts, including animism, survived from around 4000 BCE with the rise of the Sumerians to the mid-19th century with Pasteur's famous spontaneous generation experiment [1,2].

Louis Pasteur showed that life as a phenomenon is not a result of ethereal interactions between inanimate matter - after brewing a nutrient-rich broth and bringing it to boil in an "S" shaped glass flask (the process now called pasteurisation) he showed that life can only start from other life [3,4]. The trait of being alive is hereditary, not self-generated. Pasteur's observations resulted in a new dilemma: if life is a matter of inheritance, then how can it have a true beginning? These results demanded a different explanation to a problem which had been and continues to be rooted in dogma. About the same time, in 1859, Charles R. Darwin published the first edition of his book "On the Origins of Species'", shedding some light upon the consequences of inherited traits and setting the stage for evolutionary biology [5]. Darwin identified a mechanism for speciation and evolution through natural selection, but he avoided a serious explanation for life's origins [6]. In many ways, these two masterpieces of scientific inquiry paved the way to the pragmatic and scientific approach to life's origins that is used today.

Contemporary approaches to explaining life's origins aim to show how a process that was impossible in Pasteur's sterilised flask, is possible on sterile planetary bodies, often invoking the mechanisms for diversity identified by Darwin. But research into the origin(s) of life (OoL) has never constituted a discipline in its own right, and instead, it borrows technical advances and insights from a variety of specialist fields. Communication is often hindered by the diversity of fields represented, each of which brings their own technical and methodological approaches [7]. In a previous paper authored by some in our community, we articulated conceptual heterogeneity in OoL research [8]. Here, we articulate the methodological heterogeneity in the field—via experiments, models, and simulations– to help realise the goal of cross-pollinating knowledge among specialist fields within the community.

To help advance a collective understanding of the OoL problem and the contemporary state of knowledge, this article reviews the basic methodology used by different scientists working on the problem. Topics are split into three broad categories: 1) experimental techniques including analysis of small molecules, materials, and sequencing of biopolymers, 2) database and data-driven computational resources, and 3) theoretical and modelling tools from quantum chemistry and thermodynamics to network methods, and phylogenetics. Each of these three topics is incredibly diverse in its own right, and might never be discussed together in another context. We embraced this heterogeneity to highlight the diversity of work required to understand the OoL and to illustrate how these can mutually inform each other. As portrayed in Figure 1, the goal of this review is to present the methodologies and techniques commonly used in OoL, rather than to be an in-depth review of any idea in particular, or a synthesis of the current questions or research paradigms in the field (which at this stage can only be a fractured view of distant ideas). For each technique we present basic introductory details and highlight a few examples relevant to OoL research. We anticipate many readers will find content in their area of expertise simplistic– this is the goal, to communicate the basics with interested scientists so they can expand their operational knowledge within the field. Sufficient citations are provided to guide the reader towards more depth wherever their interest may lie. We hope that our work is both an educational tool and an inspiration for future post-disciplinary collaborations in OoL research, by helping scientists understand *what* they can do about the problem of life's origins, rather than telling them *how* to think about it.

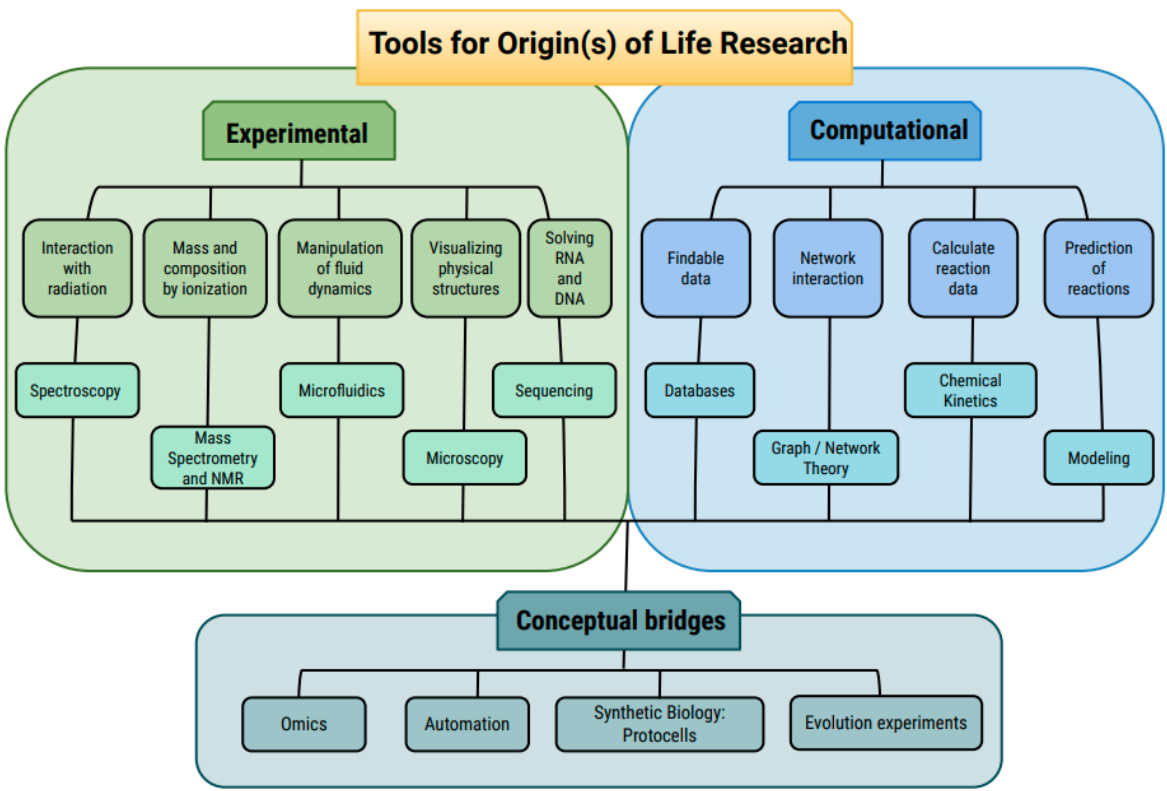

**Figure 1.** Comprehensive array of experimental and computational techniques, along with conceptual bridges, which are primarily utilised in OoL studies.

# 2. Experimental techniques for studying the Origin of Life

Life on Earth manifests in chemical substrates, and accordingly many approaches to understanding the OoL aim to analyse different molecules in the laboratory. Different disciplines employ different analytical techniques depending on the subject studied. For example, when requiring information about exact molecular structures of a small subset of pure compounds, synthetic chemists tend to use spectroscopic and separation methods focused on specific chemical targets. Geobiologists and geochemists tend to use techniques characterising bulk elemental ratios, or diversity of compounds in complex media. Molecular biologists may be more interested in the molecular sequence of a large molecule that is already identified as a peptide or RNA strand. These different approaches are rarely considered together, but understanding the Origin(s) of Life (OoL) will require understanding phenomena across and between these scales. Therefore, in this section we review the basics of analytical techniques most used in OoL research, including the physical-chemical principles they employ, and their strengths or limitations (Figure 2).

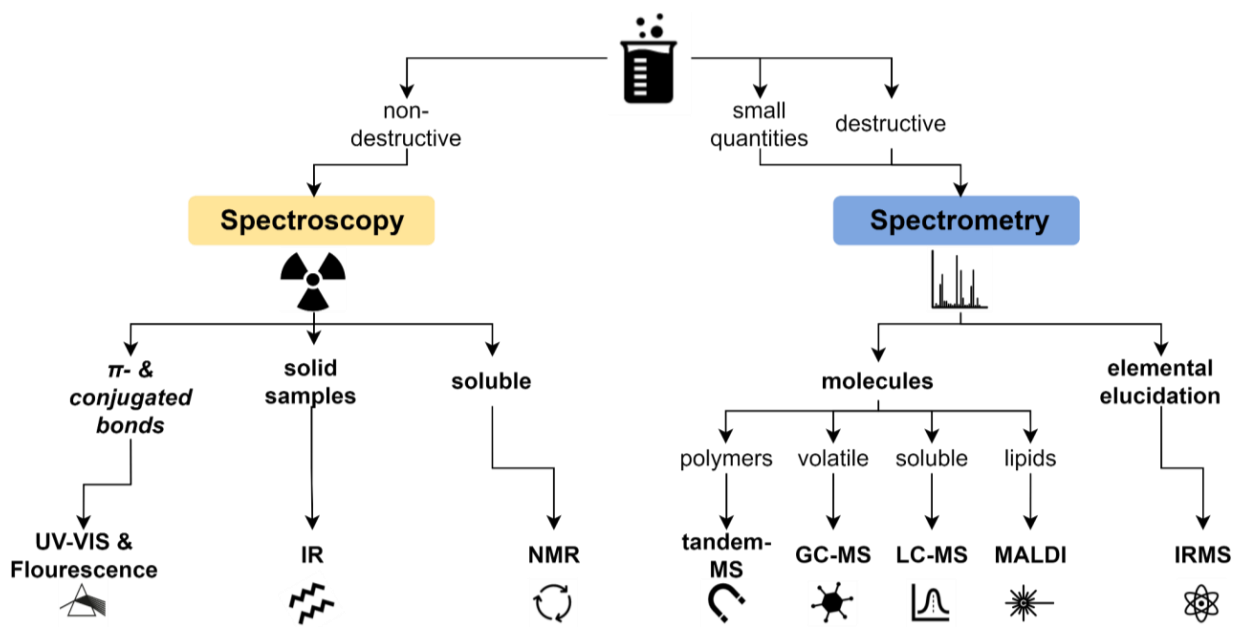

**Figure 2.** Summary of the OoL analytical techniques targeting different compositions separated into two major categories: Spectroscopy and Spectrometry described individually in the Section 2.

## 2.1 Spectroscopy

Spectroscopy deals with the interaction of matter with radiation and the resulting spectra it produces. The type of radiation source, as well as the way materials transfer the radioactive energy (adsorption, emission, scattering, photo- or chemiluminescence, etc.), will determine the spectroscopic technique. As a non-destructive technique [9], spectroscopy allows the analyses of samples without disruptions in the molecular environment, which is advantageous in the study of rare samples. In principle, a single sample could be analysed by different spectroscopic techniques, permitting the identification and quantification of individual molecules in liquid, solid, or gaseous samples with little to no sample preparation required.

High coverage of the electromagnetic field translates to precise identification of molecules. Because of its versatility, spectroscopy has been one of the most important tools for molecular identification in chemistry for decades [9–11]. In an OoL context, visible (Vis), ultraviolet (UV), infrared (IR), Raman, and nuclear magnetic resonance (NMR) are commonly used spectroscopic tools [12–14]. Here we summarise the principles behind these spectroscopic techniques used in prebiotic chemistry and how to extract meaningful information for a comprehensive understanding of different sample types.

### 2.1.1 Ultraviolet-visible & Fluorescence Spectroscopies

Ultraviolet-visible (UV-Vis) light spectroscopy is a simple and inexpensive analytical procedure that is widely exploited in the fields of analytical chemistry and biotechnology. The technique relies on the absorbance of light in the UV-Vis light range (~100 - 750 nm), where most molecules and ions absorb with different relative intensities somewhere in this range [15]. The absorbance spectrum of species can serve as a diagnostic test for their presence in solutions and can be used to relate the quantity of absorbance with the concentration when the absorbance coefficient is known [16]. Fluorescence spectroscopy is a related technique which monitors the emittance of light at a separate wavelength following absorbance. This technique has a greater level of sensitivity than UV-Vis and tends to be more specific with an order of magnitude of difference for detecting and quantifying the intensity of metabolic or fluorescent biomolecules. Both techniques are typically employed in aqueous solutions but gas phase techniques are possible [17]. Detection of individual compounds in complex mixtures can be achieved with these techniques by coupling analysis to chromatographic techniques such as HPLC [15].

In OoL research, UV-Vis and fluorescence spectroscopy are commonly employed in small molecules analysis. Many analytical assays utilize the formation of a UV absorbent or fluorescent molecule as a means of quantitative determination, examples with relevance to OoL research include the formation of acetyl phosphate [18], the determination of free thiols [19], and carbamylated amino acids [20] to name a few. Derivatization (a chemical reaction to make certain analytes measurable by changing the functional group [21]) procedures are often necessary to conjugate strongly UV absorbent or fluorescent groups to weakly absorbing molecules which helps to improve chromatographic analyses (for example sugars [22] and amino acids [23]). Nucleotide species absorb well at 254 nm and have been detected using HPLC-UV [24,25]. UV-Vis has been used to monitor the formation of inorganic iron sulphur clusters [26,27] and small molecule interactions in solution [28].

Circular dichroism spectroscopy (CD) is a type of spectroscopy that leverages the fact that materials can differentially absorb light of different polarizations. OoL studies have used this technique to understand the unique self-assembly behaviour of guanosine monophosphate nucleotides [29]; for evaluating the secondary structures of peptides and their significance in early Earth peptide chemistry [30,31]; the role of transition metals in prebiotic oligomerisation of depsipeptides [32]; and the critical role of pH in non-thermal RNA strand separation and hybridisation in the context of early Earth conditions [33]. More recently gas phase CD has been applied to explain the plausible role of gaseous phase amino acids and their photo reactivity in selection of L-amino acids in contemporary life forms [34].

### 2.1.2 Infrared Spectroscopy

Infrared (IR) spectroscopy relies on energy absorbances that range from 900 nm to 1 mm. Studies analyse different IR wavelength ranges according to the different applications and searches: 1) 30-1000 µm for rotational spectroscopy (far-IR), 2) 1.4-30 µm for fundamental vibrations and rovibrational structures (mid-IR), and 3) 0.75-1.4 µm for harmonic vibrations (near-IR) with unique spectral properties according to the elements and molecular species analysed. Identification of organic compounds can often be accomplished by mid-IR where the radiation source can be a Nernst glower, globar, or laser.

IR observations can characterise geological matrices and determine the presence of certain minerals (*e.g.* olivine) [35] or biogeological formations (*e.g.* within stromatolites) [36]. Additionally, IR can characterise organic molecules in chemical standards, biological cultures, and environmental samples [37]. IR can also be a powerful complementary technique to compare and study mostly qualitatively different samples. For instance, in the case of spaceflight analyses on primitive bodies (*e.g.* comets, asteroids), IR analyses characterise prebiotically-relevant molecules (*e.g.* water, polyoxymethylene) [38–40]. Spaceflight satellites carry on generally an IR (or UV) spectrometer that helps identify chemical families and conduct kinetic studies or look for (bio)chemical-markers of habitability [41] or life-search [42] on a planetary system. For example the search of water, methane, carbon monoxide with the Mars (*e.g.* Viking Missions), small bodies (*e.g.* Rosetta) probes [43,44], and exoplanets [45,46], organic compounds in rich diversity like on Titan with the Cassini orbiter [47–49]. IR observations are usually validated by other analytical techniques such as mass spectrometry, which will be detailed later on (Section 2.2) [50]. In OoL research, scientists have used IR coupled with X-ray surface analyses to study both synthesised processes in a laboratory and environmental historic samples (from 3.7 to 4.2 billion years ago) in the field for possible molecules and organisms sheltered or produced [51,52].

### 2.1.3 Nuclear magnetic resonance spectroscopy

Nuclear magnetic resonance (**NMR**) spectroscopy uses changes in the local magnetic field of the atomic nuclei of interest [53]. NMR relies on the detection of elements with an odd number of protons or neutrons, e.g. $^1$H, $^{13}$C, $^{15}$N, $^{31}$P etc. The location and intensity of the peak in NMR will depend on the resonance frequency of the nucleus, the environment around the nucleus of interest and the strength of the magnetic field of the instrument. NMR is used as a qualitative and quantitative technique.

Due to its sensitivity and versatility, NMR is widely used in the context of OoL studies. For instance, it has been used to verify the synthesis of organics from simple compounds in a wide range of concentrations and conditions [54,55]. In addition, NMR has been used to investigate the formation of iron-sulphur clusters in prebiotic conditions [26,27]. By equipping with a variable temperature unit, *in situ* analysis (average spectra) can be used for example for the quantification of the ribose sugar conformations in solution at different temperatures [56]. To understand the interaction of organic molecules with Si on early Earth, a combination of $^1$H and $^{13}$C NMR with $^{29}$Si NMR was used, demonstrating the interaction of sugars in silicate solution [57]. This could be coupled with the distortionless enhancement by polarisation transfer (**DEPT**) technique of $^{13}$C NMR, which is based on the variation of pulse sequences and the NOE (nuclear overhauser effect), to determine the exact type of carbon ($CH_3$, $CH_2$, CH) that is interacting with the nucleus of Si. $^{31}$P NMR is widely used in the field to study phosphorylation [58,59], nucleotide activation chemistry [60], polymerisation of mononucleotides [61], and to characterise the type of phosphates (and P-linkages) in oligomers [62].

Two-dimensional (**2D**) NMR yields 2D data in a space defined by two frequencies rather than one. Types of 2D NMR include correlation spectroscopy (COSY), J-spectroscopy, exchange spectroscopy (EXSY), and nuclear Overhauser effect spectroscopy (NOESY). For example, diffusion-ordered spectroscopy (DOSY) NMR is used in the determination of the diffusion coefficients of molecules under a given magnetic field [63]. This can be used, for example, to detect the presence of larger molecules under polymerisation conditions in comparison to the diffusion of the starting monomers.

## 2.2 Mass Spectrometry

The principle behind mass spectrometry (**MS**) is to use the motion of charged particles through known electromagnetic fields to learn about the mass, composition, and sometimes structure of those particles [64].

MS instruments ionise the sample and measure the mass-to-charge (m/z) ratio of the generated ions. From the m/z ratio, the mass can be calculated and used to search for targeted molecules or to identify general molecular formulas. The vast amount of different ionisation techniques allows for a variety of materials to be analysed, and MS is even used in the context of space exploration [65]. The high sensitivity of the instruments makes MS a versatile tool to analyse pure molecules and mixtures and allows for the analysis of small sample quantities. However, the high sensitivity is the reason for its main disadvantages: the vast amount of data collected in each measurement, the instruments' proneness for contamination, and false positive detections [66]. Mass spectrometry helps characterise molecules or molecular complexes with one order of magnitude lower than NMR and allow an absolute-quantification compared to relative or semi-quantification performed in Spectroscopic, XRD, or Raman analyses.

## 2.2.1. Molecular analysis

MS instruments come in many different varieties, but the main differences between all types of MS are the ionisation and the detection method. The ionisation method describes how the molecules are ionised and different methods will enable the detection of different types of compounds in different situations. For example, electrospray ionisation applies a high voltage to the liquid sample as it flows out of a capillary, atomizing the sample into tiny charged droplets. These will split into charged ions as the solvent evaporates, thus allowing the analytes to enter the gas phase. The detection method (or mass analyser + detector) describes how the ionised compounds are separated by their m/z ratio and then detected by the instrument. The detection method will determine the mass range, and resolution of measurements made using the technique. For example, in time-of-flight MS (ToF-MS) the m/z ratio is determined by measuring the velocity of ions that are accelerated when an electric field is applied to them as they are generated by an ion source. A detailed understanding of an MS analysis requires understanding the details of both the ion source and the detection method.

Matrix-assisted laser desorption ionisation (MALDI) is often used with ToF-MS. Here, the sample is required to be uniformly mixed in a matrix which absorbs the energy of the laser and converts it to heat energy so that the sample is not fractionated. The rapid heating allows for a small part of the matrix to be vaporised together with the sample, generating charged ions of various sizes. This technique has been used in OoL research to detect the polymerisation of RNA [67,68].

Laser desorption techniques (such as LD-MS, INMS, and MALDI) have been used to quantitatively detect biosignatures on extra-terrestrial surfaces and primary bodies (*e.g.* comets, asteroids/meteorites) without requiring a matrix to analyse the organics [69,70]. MS techniques have been employed in OoL studies, such as the determination of molecular information from biotic origin in minerals (bio monomers have been selected for life and assembled in biopolymers on mineral matrices) [71], proto-ribosome formation to produce proteins [72], and phosphoribosyl pyrophosphate to produce nucleotides as AMP on a mineral matrix [73]. These metabolomic studies help understand how building blocks of life have been produced on early Earth.

In many situations, multiple analytical techniques can be applied in sequence or simultaneously (and then their abbreviations are hyphenated) to integrate signals from different instruments in order to improve detection and quantification of products for an entire bulk sample. This allows the limitations of individual techniques to be overcome with complementary approaches, and it is particularly common with MS. When coupled to a separation step (*e.g.* a chromatographic technique) MS can elucidate the identity of many compounds in complex mixtures [21,74–76]. These chromatographic analyses use a suitable column to separate organic molecules based on the affinity of the targeted molecules to the surface of the column. The separated molecules are then directly injected to the MS. In addition to having analytes in the gas phase, liquid samples can also be analysed using a liquid chromatograph. High-performance liquid chromatography (HPLC), conventionally coupled to detectors such as UV, or Fluorescence spectroscopy (see Section 2.1.1), can also be coupled to MS, allowing the quantification of analyte concentration in solution.

The most versatile applied MS techniques relevant to OoL studies are triple or quadrupole MS, such as GC-MS and HPLC-MS, and derivative techniques [77]. Denaturing or partially denaturing by GC/LC-MS is recommended for the detection of single nucleotide substitutions and mutations of DNA/RNA sequence [77], detection of short nucleic acid strands formed in abiotic polymerisation reactions [61] and for templated primer copying [78]. The combination of HPLC-MS with a UV detector is used extensively to study RNA and DNA oligomers. A recent study presents a similar method for the accurate analysis of formose reaction products, a known combinatorial explosion that entails hundreds of compounds in solution [79].

Another MS technique to identify molecules is tandem mass spectrometry (MS/MS or MS$^n$), in which the analytes are fragmented multiple times [64]. In the first mass-analysis step (which is identical to conventional MS) parent species are identified. These parents are then fragmented and reanalysed by a second MS. In many cases, this process can be repeated several times (MS$^n$) but since the molecule is fragmented in each step leading to more fragments and fewer ions overall, more than two fragmentation steps are uncommon. This technique has been used in recent studies to assign the complexity of biological molecules [80] or for sequencing non-conventional oligomers which have been used to study the origin of functional peptides [81].

## 2.2.2. Elementary analyses

While MS is very commonly used for the identification of molecules in sample mixtures, it can also be used for elemental elucidation. Isotope-ratio MS (IR-MS) analyses the elemental content of a sample with discrimination made between the isotopic forms such as $^2$H/$^1$H, $^{13}$C/$^{12}$C, $^{15}$N/$^{14}$N and $^{18}$O/$^{16}$O, and the organic content [82]. The determination of the relative abundance of a non-radiogenic element provides information about isotopic fractionation in a sample due to the origin of the material (segregation of carbon and nitrogen heavy elements towards the light in biological systems, for instance) and the environment (we can retrace the paleoclimate of an area through the segregation in hydrogen and oxygen present in liquid/gas/icy water, for instance). The determination of the radiogenic relative abundance is generally used to date or trace the material in the different strata of a sediment/icy core (e.g. $^{11}$C/$^{12}$C, $^{40}$Ar/$^{39}$Ar, $^{87}$Sr/$^{86}$Sr, $^{206}$Pb/$^{204}$Pb) [83]. Isotopic fractionation can also be used to segregate matter coming from different celestial bodies (comets, asteroids, planets, natural satellites, etc.) to aid in the determination of the elements present in a protoplanetary disk (presolar nebula) and the early Earth, thus informing us on the prebiotic bulk necessary for the formation of the building blocks of life [84]. IR-MS and associated isotopic techniques are used to resolve questions on the most ancient trace of life on Earth in microbialites to return to the Last Universal Common Ancestor (LUCA) and First Universal Common Ancestor (FUCA) organisms [85]. *In situ* isotopic analysis is also particularly useful for determining the origin of organic matter. On the Martian soil, the SAM (Sample Analysis at Mars) experiment (aboard the Curiosity rover) has allowed after-combustion experiments (>550°C) and a comparative study of the isotopic carbon distribution between Martian samples analysed by SAM and analyses of Martian meteorites, and concluded that endogenous organic matter had indeed been detected [86].

Inductively coupled plasma MS (ICP-MS) ionises and analyses a sample to produce atomic (or small molecular) ions. Compared to atomic absorption spectroscopy, ICP-MS has higher speed, precision, and sensitivity on a larger range of atoms, which helps for detecting at trace amounts (down to a few parts per trillion) different elements and their isotopes. The advantage of the technique is the analysis of a liquid sample as well as solid samples that have to follow an acidic digestion (lixiviation) in a mineralizer (e.g. microwave mineralizer). The coupling of the ICP with mass spectrometry allows for very efficient ionisation of the analysed substance and therefore for very high sensitivity. This permits analyses at very low concentrations, including trace elements in geological samples [87]. This quantitative method is also useful to calculate isotopic relative abundances, or to quantify each element on a wider range of atoms than other instruments, for instance [67,88]. As for IR-MS, this technique is useful through the isotopic composition, to trace the origin of the elements in meteorites [89], for instance, and determine the stress or the type of chemical evolutions elements pass through in time with the analyses of succeeding strata of meteorite/sediment/icy cores.

## 2.3 Microfluidics

Obtaining reproducible and detailed results has often proved difficult for OoL studies aiming to simulate complex geochemical scenarios often involving heterogeneous phases and out-of-equilibrium dynamics. For instance, using a large (2L) open-flow reactor simulating silica-rich hydrothermal percolations on the Hadean ocean floor, Herschy et al. [90] reported the non-enzymatic synthesis of high-energy carbon reduction products such as formaldehyde in µM concentrations from $CO_2$ and $H_2$. However, these results were difficult to reproduce systematically, most likely due to the large stochasticity involved in both the manual sampling of the precipitates and the 3-dimensional arrangement of these with respect to the soluble reagents.

Microfluidics, the control and manipulation of fluids constrained to a small scale (often µL) where surface forces prevail over volumetric ones, has emerged as an alternative approach. A strict control of the experimental variables can be achieved using this technique, thus yielding robust and reproducible results [91,92]. On top of this, obtaining a laminar flow regime - in opposition to a turbulent one - for the liquids used in microfluidic experiments allows for establishing constant out-of-equilibrium conditions. This way, two solutions of differing compositions can be made to flow in contact with one another without mixing - and thus the immediate dissipation of the associated free energy. For example, [93] demonstrated how useful microfluidics can be for generating steep and stable pH gradients over µm distances. Further origin-of-life studies have focused on how these gradients could be non-enzymatically tapped in order to fuel autotrophic prebiotic chemistry scenarios [94]. Microfluidics is not only useful for controlling the fluid dynamics of an experiment, but modifications in the chip's architecture allow for precise and *in situ* measurements of reaction parameters such as voltage, temperature, or pH [95].

## 2.4 Microscopy Techniques

In some cases, assemblies or physical structures containing chemicals are large enough to be visualised directly by imaging techniques, which often utilise microscopy techniques. A significant number of technological advances have been made in the efficacy and resolution of microscopy techniques, resulting in the widespread use of microscopy in a variety of applications which require information about the physical structure of an object which is not visible to the naked eye. Some microscopy techniques can be combined with spectroscopy, adding a layer of chemical information to the information each microscopy type has to offer. Most often Raman and IR are used to achieve this.

Such microscopy techniques have also been applied to OoL research [96], especially as a mechanism to visualise supramolecular structures assembled from primitive chemicals or chemical ensembles, including protocells and other primitive compartments [96–100], nanofibers and prebiotically relevant molecular self-assemblies [101–103], and mineral structures (both terrestrial [104] and extra-terrestrial [105]). Here, rather than providing a detailed review on the entire breadth of microscopy techniques available to researchers, we briefly highlight a selection of microscopy techniques commonly used in OoL research.

### 2.4.1 Light and Fluorescence Microscopy

Light microscopy (also known as brightfield microscopy) passes light through a sample (usually prepared on a transparent surface). The sample affects the light that passes through it by absorption, scattering, or deflection, before passing through an objective (to magnify the object), ultimately reaching a detector or a camera. The pattern that reaches this detector results in the image that is acquired [106], and images acquired over time can be strung together to form a 'movie' of a sample. However, depending on the physical structure of the sample it is not always possible to distinguish the patterns that passed through the sample than one that did not. In this case, other methods taking advantage of the properties of light itself (such as phase shifts of the waves or changes in the polarity of the incident light after interaction with the sample) are used to image the sample, and observe for instance chiral compounds (associated to life which favoured an enantiomer over the other one). These include phase contrast microscopy [107], differential interference contrast microscopy (also known as DIC) [108], and polarisation microscopy [109], each of which has been applied to OoL studies for both static image and movie acquisition (*e.g.*, [110], [111], and [112], respectively).

In some cases, it is not possible to visualise samples by direct light, such as due to their lack of contrast with the surroundings. In these cases, fluorescence signals can also be detected in samples in a technique termed fluorescence microscopy [113]. In fluorescence microscopy, the light at a specific excitation wavelength is applied to a sample, and fluorescent molecules within the sample then in turn emit another specific wavelength of light, the pattern of which is passed through an objective and observed by the detector which results in an image. Some chemicals and materials are naturally fluorescent, such as certain protocells [114], lipids and proteins [115], RNA and DNA [116], and other organic compounds useful for biogeochemistry studies [117], in addition to fluorescent mineral studies [118]. The environmental or synthesised samples for OoL studies containing these natural fluorescence sources could be imaged directly (or at least with slight processing/sample preparation) on a fluorescence microscope. However, samples without such natural fluorescence often need a 'fluorescent tag' to be added into the system and visualised by fluorescence microscope. These 'tags' can be small molecules or proteins which can be non-covalently incorporated into the sample (e.g. fluorescent thioflavin T labelling of peptide and DNA [119], or fluorescent SYBR Gold intercalation into nucleic acid duplexes to identify species or quantify potential prebiotic oligoribonucleotides [120,121]). Another option to fluorescently label a sample is to covalently link a fluorescent molecule to a component of the sample so that when the sample is produced, the fluorescence is already intrinsically apparent within the sample; examples of this include green fluorescent protein (GFP) fusions [122,123] or fluorescein labelling of nucleic acids [124]. Such fluorescence microscopy techniques can also take advantage of fluorescence transfer of pairs of fluorescent molecules (*i.e.*, the fluorescence emission wavelength of one molecule equating to the excitation wavelength of another) to visualise molecular interactions between fluorescently tagged components in a sample through fluorescence resonance energy transfer (FRET) microscopy [125,126]. Single-molecule particle tracking methods using fluorescence microscopy have also been used to analyse the diffusion [127] and coalescence/wetting [128] properties of membraneless protocells.

### 2.4.2 Confocal Microscopy and Optical Coherence Tomography

While large amounts of direct incident light for fluorescence excitation (*i.e.*, epifluorescence) is useful for general imaging of samples by microscopy, in some cases, it is necessary to glean more spatial information from a sample. In particular, direct observation on an epifluorescent microscope relies on a large amount of excitation, which maximises the emission of the signal of the sample at the cost of spatial resolution in the z-direction (*i.e.*, height of the sample). While for some flat samples, this does not come at much of a cost in resolution, for larger bulkier samples, z-direction resolution could be significantly lost. Thus, one modern fluorescence microscopy technique that is used to obtain very detailed z-direction resolution is confocal microscopy [129,130], employed for the giant vesicles 'colonies' hypotheses [131] or the study of organic-mineral interactions [132], for instance. Similar to epifluorescence microscopy, confocal microscopy uses incident light to elicit a fluorescence excitation response in a sample. However, a confocal microscope uses a laser to excite a specific area of the sample, controllable in the x-/y- plane of the sample (*i.e.*, the width and length of the sample). This way, the damage to the sample is minimised - photobleaching generally occurs over long periods of exposure to the excitation light, which may result in sample destruction or destruction of the fluorescence signal [133,134]. Additionally, by using a feature called a pinhole, only the emitted fluorescence signal that is completely in focus is detected, and a pinhole of a sufficiently small size will allow the detector to acquire high-resolution images of a very thin 'slice' (in the z-direction) of the sample. By acquiring such 'slices' at different heights of a sample, a very high-resolution 3-D reconstruction of a sample can be obtained.

Confocal microscopy also gives the ability to perform analyses which allow researchers to understand more about the complex dynamics of a system. One such technique is fluorescence recovery after photobleaching (**FRAP**) [135], where the excitation laser in a confocal microscope is powered up to purposely bleach a specific area of a fluorescent sample. Then, the area is monitored for increases in fluorescence signal over time, with the rate of the fluorescent recovery determining physical properties of the system, such as the exchange rate of molecules into/out of a compartment [136], the diffusion constant of the fluorescent molecules [137], or even the phase properties of the sample (*i.e.*, whether a sample is solid or liquid) [138].

### 2.4.3 Electron Microscopy

Electron microscopy (**EM**), uses a beam of accelerated electrons to produce an image. The resolution is significantly higher than that of light microscopy due to the short wavelength of electrons. Depending on the type of electron microscope and the mode it is operated in, this technique can reach pm range resolution. There are various applications for the different types of electron microscopy, so this section will focus on the geological applications in an OoL context and will show possibilities for extending these techniques to other disciplines.

There are two main types of EM – scanning (**SEM**) and transmission (**TEM**). In geological applications both are often used to provide key information on the detailed structure and elementary distribution of minerals and organic compounds that could be related OoL [139–143]. SEM produces images by scanning the specimen with a focused beam of electrons and analysing the energy lost by the electron-specimen interaction [144]. TEM requires a beam of electrons to go through and interact with the specimen to produce a high-resolution image [145]. It is used to view the nano to atomic scale interior of the specimen, such as the organisation of organic molecules, the nano-structure of inorganic minerals, as well as elemental distributions to retrace the organic origin or date rocks and fossils even if often debated [146–148]. For TEM analysis, the specimen must be prepared as an ultrathin (less than 100 nm thick) foil so that electrons can pass through generating a projection image. This technique can also be used for (nano)particulate matter, using small grid plates to position the specimen in the electron beam. On one hand, **TEM** can generally produce higher image resolution than **SEM**. On the other hand, as **SEM** images the surface of the specimen rather than its interior, it does not require extensive preparation and enables the study of very pristine samples. Both techniques are usually combined with elemental analysis (**EDS**), adding further information.

Recent studies report various associations between catalytically active minerals and surrounding abiotic carbonaceous matter in the oceanic lithosphere (which can provide free energy for synthesis [149,150]) using **SEM** or **TEM** combined with the spectroscopy [142,143,151]. By providing high resolution structural imaging of the rock, these observations highlight the catalytically active nanocrystalline minerals, and nanoconfinement effects of the porosity for abiotic organic synthesis.

Based on the geological applications in an OoL context, it will be critical to evaluate the surface chemistry of catalytically active minerals and rock structures. With *in situ* liquid and/or gas-cell TEM and atomic-resolved aberration-corrected TEM becoming readily available [152,153], the identification of surficial catalytic sites directly during prebiotic hydrothermal experiments is possible. These new generation electron microscopy techniques will allow us to obtain unprecedented new knowledge about the physicochemical mechanisms operating during prebiotic organic synthesis.

## 2.5 Genomic Sequencing

Nucleic acids are ubiquitous in modern biology, and it has been postulated that primitive systems on early Earth also incorporated nucleic acids [154] or nucleic acid-like molecules [155]. Whether nucleic acids were necessarily required for the emergence of early life and, if so, what form these nucleic acids took on—*i.e.*, whether RNA or some other form [156,157] is still under debate and has been covered at length elsewhere [158,159]. Nevertheless, nucleic acids are present in all of modern biology, which means that they have existed at the time of the Last Universal Common Ancestor (LUCA) [160]. While previously mentioned techniques can reveal a significant amount of information regarding nucleic acids, identifying the sequence of nucleic acids directly (*i.e. sequencing*) remains an important aspect of both primitive and modern nucleic acids research [161].

### 2.5.1 Sanger Sequencing

Sequencing of nucleic acids has been performed since the 1970s with the invention of Sanger sequencing, which utilises a combination of enzymatic polymerization and gel electrophoresis to sequence a sample [162,163]. In OoL research, Sanger sequencing is a commonly used technique to confirm and analyse genetically-modified prokaryotes. Briefly, a high concentration (*i.e.*, many copies) of a sequence to be analysed is subjected to a standard primer extension reaction, with the sequence to be analysed acting as the template. The products will then be subjected to gel electrophoresis, where products will be separated by charge and size. Because each product differs in length from the next by one single base, the polymerization products will form a 'ladder' on the gel, which represent products terminated at each base by incorporation of the non-canonical monomers. The distance between each ladder component on the gel (representing each termination product) indicates the size of the base that was incorporated because each of the canonical bases has a different molecular weight, and so by cataloguing the distance between each ladder band starting from the bottom of the gel, the sequence of the analyte can be built. With particular interest for OoL research, the reading of polymerized sequences on gels (akin to the Sanger sequencing process) has been used to analyse incorporation, fidelity, and efficacy of non-enzymatic single or multiple base additions (using activated nucleotides) to a primer-directed polymerizing RNA template [78,164], revealing additional mechanisms by which primitive RNA molecules could have replicated in the absence of enzymes.

### 2.5.2 Next Generation Sequencing

While Sanger sequencing is the 'Gold Standard' of nucleic acid sequencing, its low throughput (about 800 bases per run, which may take a few hours) limits its use to a few very specific applications. The so-called 'next-generation sequencing' (NGS) techniques can analyse millions or billions of bases over hours or days at lower cost [165]. Specifically, NGS allows simultaneous analysis of the nucleic acids in a highly diverse mixed sample, rather than a single nucleic acid sequence as is the case in Sanger sequencing. A number of different techniques were developed over the last two decades, including sequencing by synthesis [166], ligation [167], pyrosequencing [168], single-molecule real-time sequencing [169], and by nanopore [170]. The most utilised NGS techniques in OoL remain sequencing by synthesis and nanopore sequencing, due to combinations of cost, read-length (*i.e.*, the maximum length of sequences that can be analysed), and throughput. Additionally, both of these platforms allow not only the analysis of DNA, but also RNA (whether directly or indirectly), an important target of some OoL research.

Sequencing by synthesis, most commonly used in Illumina sequencing platforms, requires a high number of copies of each sequence, and results in a library with a significant number of copies of each sequence from the original sample. While DNA can be amplified directly by PCR, RNA must be subjected to reverse transcription to transform it into complementary DNA before amplification (*i.e.*, RT-PCR) [171]. Once these amplified libraries are prepared, they are then subjected to a flow cell where sequences attach to the surface of the flow cell. Given that a flow cell may be able to accommodate up to billions of clusters, billions of data points (*i.e.*, sequences) may be acquired in parallel, illustrating the power of throughput of this technique. However, the scale of the sequencing data of even just one run requires significant processing (spawning a significant number of new sequencing analysis techniques in the bioinformatics field [172]) to have usable, accurate data. Nevertheless, certain techniques during the library preparation phase (before sequencing), such as sample barcoding [173], can be applied to accurately quantify, categorise, separate, or increase the throughput of sequences within a library, which may decrease sequencing noise and bias.

Sequencing by synthesis techniques have been heavily used in the OoL field to study landscape profiles of evolving primitive nucleic acids that have a specific function (such as in aptamers, RNA strands which bind to small molecules, [174]), that can catalyse reactions (such as in ribozymes [175]), and that can polymerize (such as in replicating systems [176,177]). Sequencing studies on systems combining genotype with phenotype information by linking peptides with nucleic acids through display methods [178,179] have also afforded more detail into evolving functional peptide landscapes [180,181] or in synthesis and evolution of ribosomes [182]).

While sequencing by synthesis techniques have provided significant advances in understanding nucleic acid evolution and function both in and outside OoL studies, these techniques do not allow for sequencing of modified or non-canonical nucleotides, which may have been relevant on the early Earth [183]. Although mechanisms to sequence non-canonical nucleotides, such as by mass ladder analysis [184] or 'indirect' sequencing techniques requiring extra library preparation steps [185], have been used, they still require further optimization for widespread and general use and require further development to efficiently and effectively detect modified nucleotide bases [186].

One such technique that has recently been used for such purposes is nanopore sequencing [187–189] which directly sequences analyte nucleic acids (as opposed to sequencing by synthesis, which indirectly sequences nucleic acids by producing a complementary strand to the analyte) [170,190]. First, the nucleic acid to be analysed passes through a nanopore, which is composed of a protein in a synthetic polymer membrane. While passing through the pore, one by one, each of the bases elicits an ionic current. This current is unique for each base, due to their differences in electronic structure, and the sequence of the nucleic acid is inferred based on the string of different ionic currents that appear when the nucleic acid transits through the pore. Nanopore sequencing has thus been successfully applied to study the plausibility and efficacy of oligomerization of nucleic acids under simulated early Earth geological conditions, such as in hydrothermal fields [115,191,192]. Because each base elicits a unique ionic current, this means that not only can canonical bases be detected, but non-canonical bases and base modifications can also be identified based on each of their respective unique ionic currents. The nanopore system can theoretically be optimised to detect nearly any non-canonical base or base modification [193], as well as even amino acids in a peptide [194], and variations of nanopore sequencers containing inorganic nanopores or electrodes have also been developed [195]. However, given that there are hundreds of known base modifications of RNA alone [184], not to mention a very large number of non-canonical bases [157,184,196], the resolution required to distinguish ionic currents of all of these bases within the same nucleic acid polymer may be practically challenging to achieve. Such thorough analyses may only be possible through further technological development.

## 2.6 Other Analytical Techniques

The previously discussed techniques represent a significant portion of the experimental and analytical techniques used in OoL research but it is not exhaustive. Here we briefly discuss a few other techniques which have been useful in recent studies, including X-ray diffraction and Raman spectroscopy.

### 2.6.1 X-ray diffraction

X-ray diffraction (XRD) studies have been used in OoL to study reaction mechanisms, molecular structures and to determine surface interactions of organics with geological substrates. X-rays diffraction has vastly been used in determining biophysical properties of proteins and nucleic acids. A review of the incremental development of X-ray crystallography techniques and its general applications is outlined in [197]. XRD has been used to confirm the mechanism of templated non-enzymatic primer extension reaction using imidazole-activated nucleotides [198]. The authors used an analogue of the imidazolium-bridged dinucleotide and demonstrated the conformational role of the template, primer and the bridged-dinucleotide complex in aiding primer-extension. This also revealed differences between the slow kinetics of non-enzymatic ligation reactions in comparison to primer extension reactions despite the same activation mechanism. XRD has also been used to evaluate the structural aspects associated with the substitution of non-canonical nucleobases (e.g. 2-Thio-uridine in place of uridine) in short RNA strands and their base pairing propensities in RNA duplexes [199]. Further, XRD has been used to show that RNA backbone linkage heterogeneity (i.e. 2'-5' vs 3'-5' linkage) might not destabilise duplexes in a primitive RNA scenario [200,201].

Finally, XRD has been used to study the interaction between organic and inorganic materials, including in hydrothermal conditions. For example, XRD was used to characterise the formation of inorganic chemical gardens in the presence of salts and amino acids[202]. In another study, X-ray diffraction patterns showed the competition of organics and dissolved salts for the occupation of interlayer vacancies in Na-montmorillonite clays [203].

## 2.6.2 Raman Spectroscopy

Raman spectroscopy (RS) analyses vibrational, rotational, and all molecular states (compared to IR analysing rovibrational structures and harmonic vibrations) [204,205]. RS defines the minerals in a matrix and some organic functions, however, RS is more sensitive than IR or UV spectroscopy for elements that are fluorescent after a UV/IR/Visible excitation, such as Si and organics. RS probes the chemical composition in different material surfaces, especially for inorganic matter compared to IR spectroscopy where we can use Pyrex glass tubes (that absorb IR radiations) since the laser beam used in RS and the Raman-scattered light are both visible regions. Moreover, the sample preparation is easier and faster than in IR for geological/inorganic matrices (and conversely for organic matrices where a simple water evaporation is needed in IR due to water interference compared to RS where water can be used as solvent). RS can analyse a wider sample than IR because of the absence of a reference light path where we study only the scattered light. One of the advantages of Raman is to also analyse a wide range of material thanks to different light excitation (from a deep-UV, UV, UV-Visible, near-IR source) RS can as well analyse gas samples with the adequate instrumentation. Finally, for biological samples in OoL research, UV-Visible and near-IR beamlines help taking high-spatial-resolution pictures, however the intense light source might destruct the biological material after analysis [205,206] compared to IR. Thus, IR for OoL research gives an indication on the ionic character, whereas RS analyses the covalent character of the material.

In OoL, RS is used as a complementary technique to XRD, FT-IR and surface microscopy. It has found applications in studying the role of amino acids and simulated early Earth seawater mixtures in the formation of goethite [207]; the interaction of nucleobases with artificial seawater[208,209]; adsorption of amino acids on zeolites [210] etc. As seen from the above examples, RS is usually not used exclusively, but in conjunction with other techniques to establish surface reaction of organics within a prebiotic context. RS is thus useful to analyse sediments, geobiological, organo-metallic and organo-mineral complexes as well as biofilms of extremophiles mimicking potential primitive communities.

The experimental techniques outlined here have been used extensively in OoL research in recent decades. Having an operational understanding of them can help OoL scientists understand and contextualise experimental results, possible sources of error, and alternative interpretations of data. As with all science, there may be new experimental techniques which dominate the field in the coming years, but these are the primary sources of empirical data to-date. Next, we turn to understand databases of empirical analyses and their use in OoL.

## 3. Databases in OoL studies

With the advance of computational power and increasingly high-throughput data-generating methods (see Sections 2.5 and 4), multiple databases of interest to origin(s) of life (OoL) research have appeared in recent decades. To our knowledge there are no databases dedicated to solving problems in OoL, such as a database of prebiotic compounds and reactions, or ancient protein folds. This means that many of the most popular databases lack metadata and context required to unambiguously characterise key processes involved in the OoL. Accordingly, scientists must exercise caution when applying biochemical or technological data. Many databases, particularly chemical databases contain large amounts of information, which can naively suggest they contain information about most of the relevant compounds; however, the space of chemical compounds is vast and the space of chemical reactions is far larger still, even without considering the different experimental or environmental parameters.

Nevertheless, several databases have proven useful in OoL research in recent years. Here we review these databases and highlight how they have been used in the context of OoL. We've broken these into two broad categories, 1) Physical and chemical databases which contain data that may or may not be relevant to Earth's biochemistry, and 2) biochemical and biological databases which contain information relevant to Earth's biochemistry including biochemical reactions, gene sequences and protein structures. The claims made based on the analysis of large datasets can only be as accurate as the underlying data itself. Therefore, all databases here suffer from similar constraints: the data are not always perfectly curated, may be missing, incomplete, or incorrect. Many of these databases are publicly available, but not all. Some databases are proprietary and require subscription fees to access. Finally, different databases have different standards and procedures for allowing new data to be submitted. The establishment of OoL databases, with the relevant standards of reporting and metadata requirements will mark a significant transition in the maturity of OoL research.

## 3.1. Physical and Chemical Data

Access to large-scale, standardised physical and chemical data can help building theoretical models of abiogenesis, identifying candidate experimental systems for synthesising (bio)molecules and life-like molecular systems, locating missing links in research paradigms, predicting what kinds of environments may be more plausible to host OoL relevant processes, and more. Databases of chemical species and reactions can be roughly categorised into two types: databases of experimentally-confirmed reactions, and databases of rule-based (or algorithm-generated) reactions (see Section 4.3.2 for more information about reaction network generation). In this section we focus on physical-chemical data that is general and not linked to specific living organisms. For chemical data linked to living organisms, see Section 3.2 below on biochemical databases.

The location, references and other useful information of multiple databases of physical and chemical data commonly used in OoL research are summarised in Table 3.1. These databases were used in multiple tasks, such as comparison between computationally-generated libraries of molecules and databases of empirically-confirmed molecules, and detecting autocatalytic cycles among abiotic reactions.

Table 3.1. Examples of physical and chemical databases commonly used in OoL of life research.

| Database name | Website | Content and Notes | OoL Use Cases |
| --- | --- | --- | --- |
| Reaxys | reaxys.com | Chemical compounds, materials, reactions, patents, and bibliographic information (commercial). Sometimes 'reactants' and 'reagents' are not rigorously distinguished, and some reaction equations are not balanced. | [196] |
| NIST databases | kinetics.nist.gov/kinetics; webbook.nist.gov/chemistry | Chemical species and reactions (open-access). These databases are not very large, but the quality of annotations is high. | [211,212] |
| CRC Handbook | hbcp.chemnetbase.com/ | Chemical species and reactions (commercial). The quality of | |

| | | annotations is very high. | |
|---|---|---|---|
| CAS database | cas.org/cas-data/cas-reactions | Reactions (commercial). | |
| Open Reaction Database | open-reaction-database.org; | Open-access, under-construction database of chemical reactions. | |
| Radiolysis-based reaction database | doi.org/10.1038/s41598-021-81293-6 [213] | Reactions (open-access). Collected from seven decades of publications; it focuses mostly on radiolytic reactions, free radical reactions, and geochemical reactions. | [214] |

To algorithmically generate molecular structures, there are multiple software tools available, including MolGEN (closed-source) [215], OMG (open source) [216] and Surge (open source) [216,217]. Rule-based/algorithm-generated reaction databases are currently rare. Available examples include the AllChemy database [218,219] and MØD [220,221]. The AllChemy database is a partially open-access database generated by machine-learning algorithms (requires registration; some of the features mentioned in the publication [218] are not available through the online portal). MØD provides a software package for graph-based cheminformatics that can be used to generate a rule-based reaction network by specifying a set of reaction rules. To learn more about these methods see Section 4.3.2.

## 3.2. Biochemical and Biological databases

Biochemical and biological data are primarily composed of genomic sequences, the enzymes encoded in those sequences, and the molecules and reactions implicated by those enzymes. Sequence databases can contain experimentally confirmed or computationally predicted metadata. For example, in Uniprot, a star indicates if the function of a protein has been confirmed *in vitro* or just inferred computationally; the latter occurs when a protein sequence matches another in the database that has been experimentally demonstrated to have that function [222]. Other types of data include experimental standard measurements for identifying known metabolites, or enzyme structure databases to map protein sequences to plausible structures (and functions especially looking for post-transcriptional/translational modifications) [223–225].

Following extensive work on the functional annotation of genomes and linking enzymology and genomics, large-scale biochemical databases have arisen that include both biological and (bio)chemical data, for example the Kyoto Encyclopedia of Genes and Genomes [226] which has been used extensively to investigate the origins of early metabolic networks [227,228] -- for more detail on these, see Section 4.3. However, the annotations of compounds and reactions are often not detailed enough, and therefore current biochemical databases are often insufficient for detailed explorations of the biochemical space. For example, it is often unclear if a reaction is one-step or multi-step and, in some cases, a multi-step reaction and its corresponding single-step reactions all have records in the database, which creates duplicates. In other cases, the distinction between reactants, reagents, and catalysts is unclear, reactions are not mass-balanced and/or lack stoichiometric information, and different chemical species share the same name (e.g. starch, glycogen). Kinetic data are almost always absent, reaction conditions (pH, temperature, salts, buffers) are mostly missing or expressed in non-compatible formats (poor metadata), and often the phase of a chemical species involved in the reaction is not specified. In the case of rule-based/algorithm-generated datasets, the reliability of predictions about what reactions are possible may not be high, and measures of uncertainty are lacking. All of these problems are worse when we consider chemical reactions not included in modern biochemistry, many of which may be relevant for understanding prebiotic chemical processes or understanding the structure of biochemistry by comparing it to alternative possibilities.

However, advances have been made with what is available today. The KEGG database has been filtered for reactions from anaerobic prokaryotes, potentially closer to primordial metabolism [228,229]. In a similar manner, the MetaCyc database has been manually-filtered for core metabolic reactions to reconstruct a prebiotically-plausible autotrophic metabolism [230].

The location, references and other useful information of multiple databases of biological and biochemical data commonly used in OoL research are summarised in Table 3.2 These databases were used in multiple tasks, such as reconstructing ancient metabolic networks, inferring structure of ancient proteins, and exploring the origin of translation machinery.

Table 3.2 Examples of biochemical and biological databases commonly used in origin(s) of life research.

| Database name | Website | Content/Notes | OoL Use Cases |
|---|---|---|---|
| KEGG (Kyoto Encyclopedia of Genes and Genomes) | www.kegg.jp | Genomes, genes, pathways, enzymes, biochemical reactions, compounds and more. | [227,228] |
| BRENDA | www.brenda-enzymes.org | Enzymes, reactions, compounds, metadata. | [231,232] |
| NCBI (National Center for Biotechnology Information) Genome | ncbi.nlm.nih.gov/genome | Genomes; can be filtered for reference high-quality genomes (RefSeq). Includes resources, metadata, tools (e.g. BLAST) and is integrated with external databases. | [85,233] |
| JGI | genome.jgi.doe.gov | Genomes | [234] |
| RCSB/PDB | rcsb.org | Protein structures | [235,236] |
| AlphaFold | alphafold.ebi.ac.uk | Protein structures predicted by artificial intelligence. Includes an algorithm for predicting new structures. | [237] |
| Uniprot | uniprot.org | Protein sequences and associated functional information and metadata. | [238] |
| MetaCyc | metacyc.org | Metabolic pathways verified experimentally with metadata. Some functions and modules can be accessed free of charge, while others require paid subscription. | [230,239] |
| LUCApedia | eebgroups.princeton.edu/lucapedia | Organises information from different databases and publications on possible reactions, cofactors and proteins present in the Last Universal Common Ancestor (LUCA). | [227,240] |
| metaXCMS | xcmsonline.scripps.edu | HPLC-Orbitrap or any HRMS data. Used in metabolomics to reconstruct primitive metabolomic pathways, and biomolecules. | [241–244] |

# 4. Theoretical Approaches and Modelling Frameworks for the Origin of Life

To understand the origin(s) of life (OoL) we need to be able to generalise from experiments, make predictions and provide explanations. This is the role of theoretical abstractions, modelling and simulations. The types of questions scientists ask about OoL, and explanations they demand are heterogenous [8]. Accordingly, the theoretical approaches used often capture different aspects of the same phenomena. In some cases, these are first principle physical approaches such as quantum chemical models, and molecular dynamics simulations.

In other cases, they are based on the principles of biology, as is the case for molecular phylogenetics. Still in other cases we need more abstract models to understand how interactions between individual parts compose a whole (network science) or how simple rules can lead to complex outcomes (complex systems modelling). Here we outline different theoretical modelling and computational approaches to various problems in OoL literature, many of which presuppose dissimilar abstractions and are designed to answer different questions.

The recent surge in computer-driven technology has enabled the development and wide application of computer simulations, and new computational-statistical approaches to large datasets. Computer simulations allow us to formalise hypotheses, rapidly screen through various conditions, and predict and rationalise experimental observables and measurable quantities. Simulations have gained popularity across many fields of science, and are particularly applicable to the OoL research, where some processes may occur under (currently) experimentally impractical conditions, or where formal models can integrate data from multiple disciplinary sources. *In silico* experiments are also useful in examining the so-called emergent properties that stem from the interaction of several components. Both dynamical systems modelling and agent-based modelling, for instance, have shown that even very simple equations or interaction rules at the micro-scale can lead to very complicated behaviour at the macroscopic level. Meanwhile, new statistical approaches have made it possible to automatically generate hypotheses and test them using large datasets. Statistical learning approaches have been applied across many fields but are increasingly becoming useful as OoL scientists begin integrating data from different sources.

In this section we will overview different theoretical tools and modelling and simulation techniques. These do not represent theories of life's origin, but instead are methods which give us scientific insights that are not directly experimental or lab-based. The sections include molecular simulations (4.1), kinetics and thermodynamics that are involved in modelling chemical systems (4.2), network modelling (4.3), complex systems modelling (4.4), analysis based on information theory (4.5) and phylogenetics (4.6). These methods bring together knowledge of physics, chemistry and biology and allow us to apply it to the yet unknown processes at the OoL.

## 4.1 Molecular Modelling and Simulations

Molecular modelling focuses on understanding chemical-physical processes at the atomic level [245,246]. It is generally divided into two methods: quantum chemistry and molecular mechanics. **Figure 3** shows the relationship between methods, system sizes and the timescales attainable by molecular modelling methodologies. The minimal sizes of the molecular system and the timescale of an event will set the lower boundary of the practical methodology, while the importance of changes in the electronic structure will dictate the attainable system size due to computational limits. Since rare events and non-equilibrium processes are of particular interest for OoL research, enhanced sampling techniques can be used, which allow sampling a larger portion of the configuration space (see [247]).

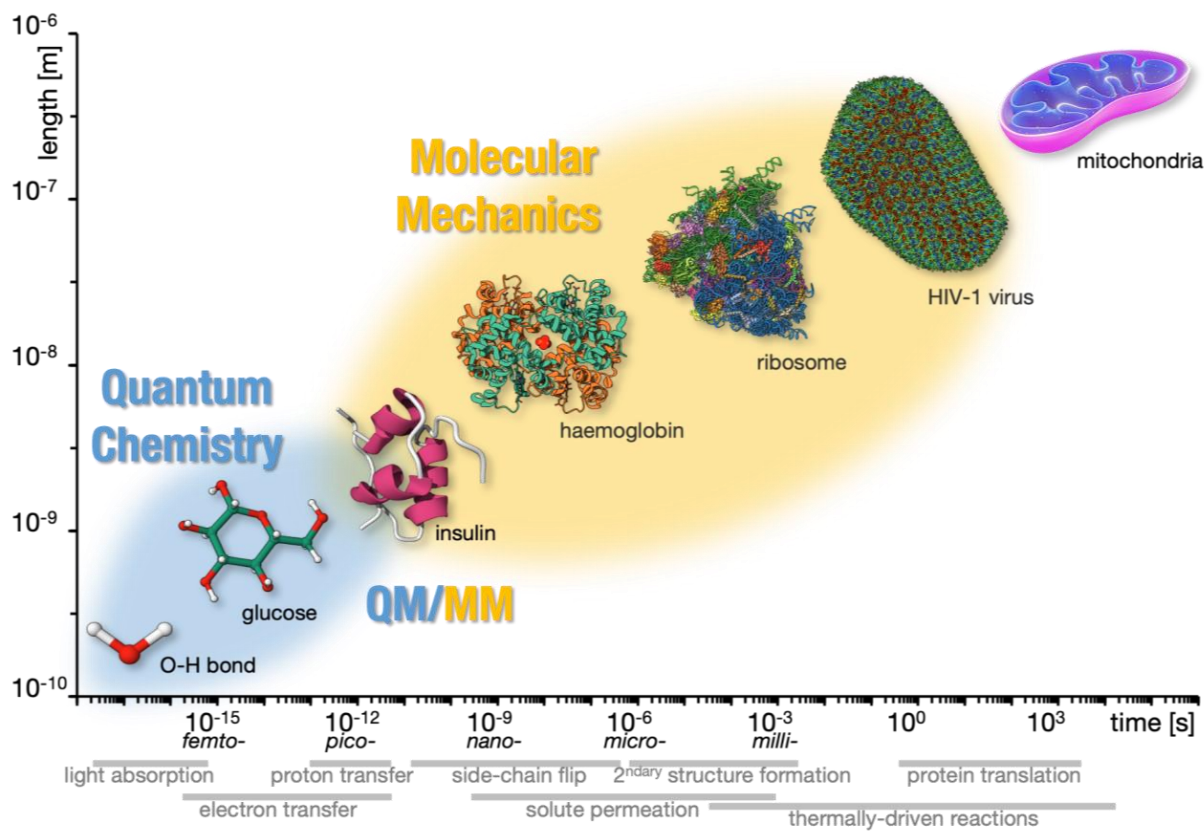

**Figure 3.** Relationship between timescale and system size in simulations from gas-phase quantum calculations (bonds, small molecules) to the condensed phase classical molecular dynamics (large molecules) and biological systems and processes (viruses, cells).

### 4.1.1 Quantum Chemistry

The goal of quantum calculations is to predict the molecular electronic structure by solving the electronic (time-independent) Schrödinger equation. Since it is only possible to analytically solve it for a hydrogen atom, systems of OoL interest require approximations to be introduced. There are many levels of approximations on the electronic Hamiltonian, which allow solving chemical structures under computational limitations. Here we summarise the approximations made and the resulting limitations of the methodology, followed by examples of their applications.

The first assumption, applicable to the majority of quantum mechanical (QM) calculations, is the Born-Oppenheimer approximation. It separates slow nuclear and fast electronic motion, which enables us to numerically find a solution to the Schrödinger equation with the electronic molecular Hamiltonian. This allows for calculation of the electronic molecular structure and energetic information for a given molecular geometry.

To this end, wavefunction-based methods describe electrons as single-particle wavefunctions (orbitals), distributed around a fixed nucleus. The early Hartree-Fock (HF) methods approximate the electronic wavefunction by the simplest possible combination of molecular orbitals (called the Slater determinant) and optimises them numerically in self-consistent-field. The main issue is that HF does not account for electron correlation effects, which contribute to chemical bonding. The development of post-HF methods, such as Møller-Plesset perturbation or coupled-cluster theory [248], have provided computational routines to account for the dynamic electron correlation. Therefore, wavefunction-based methods are typically good for calculating optimum molecular geometries, transition states, and vibrational spectra (see Section 2.1 Spectroscopy). These methods have found use in the interpretation of the extra-terrestrial spectra and identification of molecular signatures in space [249], as well as for calculating free energies of molecular interactions, such as the emergence of codons within the DNA [250]. Nevertheless, wavefunction-based methods remain applicable to small molecules only due to their high computational cost. To mitigate the issue with sampling of the molecule's configuration space, the method will often rely on the sampling done with lower-cost and lower-accuracy methods, whether quantum or even classical mechanics based.

Density functional theory (DFT), has gained popularity due to its computational affordability while delivering reliable predictions of molecular geometries and associated ground-state properties of a system. DFT describes the electron system by an electronic density instead of an electronic wavefunction – focusing on approximations that can be made on the electronic Hamiltonian [251]. Typically, the methods are based upon Kohn-Sham theory [252], which introduces an orbital representation of the electronic density to better evaluate kinetic energy, while offering a different option to approximate the (in)famous exchange, and correlation energy and potential. If this exchange and correlation potential were to be known, Kohn-Sham DFT would be exact. Unfortunately, this is not possible, so the functional form of this potential needs to be approximate, leading to a range of Density Functional Approximation (DFA) methods. DFAs are often classified in families such as Local Density Approximation [253], Generalised Gradient Approximation [254], or the very-commonly used hybrid (exact-exchange) functionals. These methods allow reliable prediction of molecular geometries and associated ground-state properties, including spectra, and allow for prediction and identification of the spectroscopic signatures of molecules applicable to their astronomical detection [255,256].

Notably, DFT can be used to study reaction mechanisms, predicting the transition states, activation barriers and pathways taken. For example, developments in the DFT allow us to revisit the hypothesis of amino acid synthesis from alpha-keto acids via catalysis by dinucleotide species [257]. The computational analysis and details of the structures for the intermediates and transition states, showed that there was wide scope for interactions between the keto acids and dinucleotide moieties and led to the required proto-metabolic selectivity [258].

While finite molecular systems are at the centre of abiogenesis research, condensed materials are also of great importance in the OoL setting [259]. Substrates (e.g., mineral surfaces, not to be confused with substrates in enzyme kinetics) may have a key role in concentrating small molecules, catalysing reactions towards the increased complexity of biological molecules [260–262]. The study of peptide bond formation on aluminosilicate surfaces allowed the investigation of this hypothesis computationally, which demonstrated the feasibility for peptide formation on the clay surface [263]. The method itself modelled substrate as an isolated cluster of atoms, assumed representative of the system. This approach represents both the molecule and substrate as a system of orbitals, which is not always applicable to materials. For example, crystalline materials or metals will have a continuum of energy levels (bands) that are not centred on nuclei, which case a planewave approach is adopted. [264]. The review of Rimola *et al* details the QM methodologies for the study of minerals with a focus of prebiotic chemistry [265].

The applications are often related to identification of potential catalytic surfaces or to provide theoretical evidence for a proposed reaction mechanism on a surface. For instance, a DFT study focused on $CO_2$ interactions with catalytic minerals such as iron sulphides (e.g. mackinawite: tetragonal FeS) revealed that when these surfaces are doped with Ni, they exhibit weaker bindings to $CO_2$ [266]. Furthermore, the DFT calculations used alongside UV-Vis (Section 2.1) experimental study have been effective in identification of products and intermediates of mineral-assisted formamide conversion to nucleobases, also allowing to identify corresponding features on the experimental spectra [267]. A QM study, coupled with an experiment and classical molecular dynamics (Method discussed in Section 4.1.2) simulations to allow for larger systems and inclusion of temperature effects, allows to elucidate interactions between organic molecules and minerals in the early Earth conditions [268].

The mechanism of formation of early organic molecules, such as formamide, has been described by **QM** models and shown to be possible at extremely low temperatures [269]. Nevertheless, not every reaction can happen in near-vacuum and at absolute zero. Therefore, it would not be fair to assume that the electronic structure of a molecule is not affected by its environment. The substrate, solvent, temperature, surrounding ions, proton gradients – are all crucial in the study of OoL. The solvent is a many-molecule matrix, coordinating to the solute molecule and creating a surrounding electrostatic medium. Due to the sheer number of atoms, the solvent is inherently computationally demanding. Polarisable Continuum (**PC**) model [270] allows to include dielectric bulk solvent effects, and by such to drop the explicit solvent molecules. Nevertheless, solvent molecules are coordinating the solute, coupling motions, allowing molecular coordination, dumping the motions and allowing for energy exchange between species [271]. Therefore, these molecules should be included. In this case, the computing effort is dedicated to the region of interest - the molecule - while the necessary but not chemically interesting areas are reduced to a minimal representation [272]. Similarly, **QM/MM** mixed methods allow the computation of a relatively large system, by reducing the surroundings of a molecule of interest to the classical (i.e. electron-free) molecular mechanics representation. For instance, to study a hydrogenation reaction of isocyanic acid on amorphous solid water to form formamide, the **QM/MM** approach allows screening the binding sites on the surface, calculating the activation energies and to identify the tunnelling mechanism of this reaction occurring at the interstellar temperatures of 103 K [273]. Similarly, a combination of a three-layer **QM/QM'/MM** framework was used to investigate hydrogen-cyanide isomerisation on an icy grain. This approach includes a larger number of molecules surrounding the reaction site and it highlights tunnelling as a main mechanism at low temperatures [274].

The temperature of the surrounding environment will result in a molecular motion, affecting properties such as molecular vibrations, anharmonic motions, and diffusion collision of species. Following BO approximation, heavy nuclear motion can be determined through Newton's law, with nuclear forces provided by calculating each time-step of the dynamics, following the gradient of the electronic energy obtained with **QM** calculations. Through the use of molecular dynamics, we can incorporate energy transfer between molecules through collisions, allowing them to overcome the energy barrier necessary for activation of a reaction. To this end, *ab initio* nanoreactor approach has been successfully used to study reactivity of aqueous HCN, suggesting that it could be a source of RNA and protein precursors [275]. Obtaining time-average properties requires a long dynamic calculation and makes *ab initio* molecular dynamics (**AIMD**) [276] an expensive method, typically allowing simulation under a nanosecond timescale [275]. To gather statistically meaningful sampling within the attainable computational resources, Devergne et al. combined AIMD with the machine learning to gather accurate free-energy profiles for prebiotic chemical reactions [277].

All of the QM methods discussed above are suitable for the ground state representation of electronic structure. While ground molecular states are applicable to many chemical scenarios on Earth, these methods cannot be used for the study of light-activated processes, such as ones occurring in interstellar space or atmospheres. These chemical processes are of a particular interest to the formation of proto-biomolecules[278]. In order to study photoexcitation, the extensions to the wavefunction methods are commonly used [279,280], and in the world of DFT, linear-response time-dependent DFT (TDDFT) is one of the most frequently used approach to study excited states of a molecule. These methods are orders of magnitude more computationally demanding, which is currently limiting their application to the study of molecules with relevance to OoL. Light has been a key component allowing for the creation of chemical complexity in the interstellar media, as in the discussed above study of hydroxylated naphthalene on dust grains, [281] or towards formation of abiotic precursors of the pyrimidine ribonucleotides [282].

This section overviewed quantum chemistry methods and their application within the realm of OoL research. Further details of this methodology, and atomistic computer simulations, applied to the study of prebiotic chemistry can be found in a review by Pérez-Villa, Pietrucci and Saitta [246].

### 4.1.2 Molecular Mechanics

Molecular mechanics (MM) is a valuable tool for modelling systems in which electronic structural changes are not of primary importance (i.e. no reactivity, excited states or changes in the chemical bonding). In these situations, molecular structures play a crucial role in determining the chemical-physical properties and functionality. The most widely used technique in MM is Molecular Dynamics (MD). In MD, classical (Newtonian) equations of motion are solved at each time step, resulting in a trajectory of atomic motions over a specified time period. Sufficient sampling is necessary to ensure that the time-averaged properties of the system are representative of the macroscopic thermodynamic ensemble. In MM, molecules are represented by atoms, each depicted as a sphere with specific radius, softness, and charge. These atoms are connected by bonds, represented by springs with specific lengths and stiffness, along with equilibrium angles and dihedrals. These parameters are defined in a force field, which is calibrated based on QM calculations and experimental measurements for specific systems. As a result, various force fields have been developed, many of which are specifically designed for liquid organic or biomolecular systems. It is worth noting that simpler mesoscale simulations, such as coarse-grained MD and dissipative particle dynamics (DPD), can be employed effectively for investigating general chemistry concepts while mitigating computational costs. Coarse-grained MD and DPD simulations are widely accepted as reliable approaches for accurately simulating chemical phenomena [284]. These methods have also been applied in OoL research [282,283].

A notable use of MD in OoL research is to provide a mechanistic explanation for the emergence of selection processes in complex mixtures. This includes selective synthesis of peptides [283] and nucleotides [283,284] on mineral surfaces, selective permeation of sugars across membranes [285] and composition-based selective self-assembly of lipids [286]. A use of reactive force fields may also facilitate covalent reactions to better study their mechanisms [287], or generate chemical reaction networks [287,288]. This theoretical framework has been successfully combined with experimental work [291], providing detailed predictions that can be explored experimentally regarding the emergence of reproduction and evolution.

Structure prediction techniques have proven useful in identifying conserved motifs and structures within the ribosome [292] and modelling short peptide sequences containing active sites of modern proteins, such as aminoacyl-tRNA synthetases. These techniques enable the confirmation of reasonable 3D structures before synthesising the sequences for wet lab experiments [293, 294]. Accelerating thermodynamic phase-space sampling is achievable through various methods that bias or modify the potential energy landscape. Examples of such methods include Metadynamics, Umbrella Sampling, and Variationally Enhanced Sampling [295–297]. Molecular docking, which allows for the assessment of molecular interactions on a large scale, provides a useful approach for quickly evaluating interactions with lower computational costs compared to quantum mechanical simulations or molecular dynamics [298]. This speed enables the exploration of problems such as the interaction of all 1280 combinations of proteinogenic amino acids with all nucleotide triplets [299]. However, its utility may be limited in systems with significant flexibility or when considering the importance of water, ions, charge, and geometry in interactions.

## 4.2 Modelling Chemical Systems

Both equilibrium and nonequilibrium approaches to modelling chemical systems are common in OoL studies. The equilibrium approach is based on thermodynamic theory, and is used in environments that locally, or globally reach thermodynamic equilibrium such as the interiors of asteroids, or the lower layers of gas giant planets. The nonequilibrium approach is based on chemical kinetic theory as well as the growing field of nonequilibrium thermodynamics, and it is used in environments that are constantly driven out of equilibrium such as terrestrial atmospheres and protoplanetary disks, and biological enzyme catalysis. Finally, many biochemical and biological systems are modelled using the principles of chemical reaction networks - which share a framework with classical chemical kinetic theory.

### 4.2.1 Thermodynamic equilibrium calculations

Thermodynamic equilibrium is the lowest free energy and most stable state of a closed chemical system. At chemical equilibrium, concentrations of molecular species no longer change with time and all (net) reaction fluxes vanish. All closed systems evolve toward thermodynamic equilibrium; however, inputs of energy and matter, such as ultraviolet (UV) light, asteroid impacts, and radioactive decay can drive open systems away from thermodynamic equilibrium. Therefore, the appropriateness for using thermodynamic equilibrium techniques depends on the system that is being modelled.

For instance, consider the atmospheres of terrestrial planets like Earth in contrast to those of gas giants like Jupiter and Saturn. Earth's atmosphere is transparent to long-wave UV light. The photon energy of UV light breaks molecular bonds and constantly forces Earth's atmosphere away from equilibrium. The molecular species ozone ($O_3$) is produced via disequilibrium UV chemistry; therefore, if one tried performing thermodynamic calculations of Earth's atmosphere, the ozone layer would not be produced. In contrast, the majority of Jupiter's and Saturn's atmospheres are opaque to UV light. Furthermore, the lower layers of these atmospheres are hot and thermodynamic equilibrium is reached for many molecular species before these gases are transported to cooler regions of the atmosphere and become kinetically inhibited [289,290].

The cooler the chemical system, the longer it takes for thermodynamic equilibrium to be established. One way to determine if a molecular species has enough time to reach thermodynamic equilibrium is to consider the time constant for the fastest reactions that produce and destroy that species. In the example above, consider a layer deep in Saturn's atmosphere, where the pressure is 10 kbar and the temperature is 2000 K. The rate-determining reaction for HCN in this environment is the HCN destruction reaction HCN + $H_2$ → $CH_4$ + NH, which has a rate constant of $k(T) = 1.08e^{-70,456/T}$ cm$^3$s$^{-1}$. The time constant is $t$(HCN,2000K) ≈ $1/k$(2000K)[$H_2$] ≈ 4 seconds, which is much shorter than the year-long timescale of convective transport in these regions. Thus, using a thermodynamic equilibrium approach is suitable for the deep layers of gas giant atmospheres.

Common thermodynamic calculations include computing A) equilibrium compositions, B) reaction abundances (yields) and C) chemical reaction affinities (i.e., whether a reaction will occur spontaneously or not). The key data for these calculations are the Gibbs free energies of formation for the species in the system, and the reaction, respectively. There are multiple databases of Gibbs free energies of formation calculated from lab measurements, including GRI-Mech 3.0 [291], CHNOSZ [292], and JANAF [293]. It is also possible to calculate Gibbs free energies of formation for specific molecules using quantum chemistry methods (e.g. [294], see Section 4.1.1).

The equilibrium composition of a system (i.e. the number of moles of each species) is that which yields the lowest total Gibbs free energy of the system. Gibbs free energy minimization solvers do not contain any information about which reactions are possible in a system, rather, they simply break down a molecular system into the corresponding atoms, and then build the system back up to the composition that minimises the Gibbs free energy. Yields for specific reactions can also be calculated by limiting the modelled system to the species involved in the reaction. There are multiple open source and proprietary codes that contain minimization of Gibbs free energy solvers, including Cantera [295], OpenCalphad [296], and ChemApp [297].

Chemical reaction affinities define whether a particular reaction is favourable to occur, and can be calculated in a few ways. Positive reaction affinities describe reactions which can proceed spontaneously. Negative reaction affinities describe thermodynamically unfavourable reactions, which can only occur when coupled to additional driving. Reaction affinities are also defined as the negative of Gibbs free energies of reaction (e.g. $A = -\Delta G$). The simplest way to calculate reaction affinities is to subtract the Gibbs free energies of formation for the products of a reaction from the Gibbs free energies of formation of the reactants, $A = \Delta G_{f,reactants} - \Delta G_{f,products}$.

Thermodynamic calculations have been performed in a variety of OoL studies, including models of nucleobase, ribose, and amino acid production in asteroid interiors [294, 299-301], lightning chemistry on the primitive Earth [302,303], impact-generated chemistry during the Hadean eon [304], Archean mantle/volcanic outgassing chemistry [305], and potential ancient metabolisms in hydrothermal systems [227,306,307].

Thermodynamic equilibrium calculations are a reasonable way to estimate the chemistry occurring in the settings above, but there are many environments that are constantly driven out of equilibrium and cannot be accurately modelled using this approach. In nonequilibrium settings, a chemical kinetics approach is often necessary.

### 4.2.2 Chemical reaction networks

Chemical reaction networks (CRN) are a general modelling framework used for studying chemical systems. It is one of the most common and important formalisms employed in OoL, biochemistry, theoretical biology and related fields. We provide a brief overview of the main concepts used for understanding CRN terminology and notation. For more details, see [308] for a gentle introduction to CRNs and chemical kinetics, [309] for foundations of chemical thermodynamics, and [310,311] for reviews of state-of-the-art nonequilibrium thermodynamics of chemical systems. A thorough review of prebiotic chemical reactions and networks can be found in [312].

Formally, a CRN consists of a set of chemical **species**, representing different types of molecules, and a set of **reactions** that convert different species into each other. Each reaction can be written in a general form as

$$\nu_1 A_1 + \nu_2 A_2 + \cdots \rightleftharpoons \nu_1' B_1 + \nu_2' B_2 + \ \ldots \tag{1}$$

This expression indicates a reaction that converts $\nu_1$ molecules of species $A_1$, $\nu_2$ molecules of $A_2$, etc. into $\nu_1'$ molecules of species $B_1$, $\nu_2'$ molecules of $B_2$, etc. The species $A_i$ consumed by the reaction are called **reactants**, while the species $B_i$ produced by the reaction are called **products**. The numbers $\nu_i$ and $\nu_i'$ specify how many reactants and products go into the reaction, and they are called the **stoichiometric coefficients** of the reaction. The overall set of species and stoichiometric coefficients that make up either the reactants or the products of any reaction in a CRN are called a **chemical complex**.

As an example, consider a situation where a biological enzyme catalyses the formation of product P from substrate S. This can be represented by the following simple CRN:

$$S + E \rightleftharpoons SE \tag{2}$$

$$SE \rightarrow P + E \tag{3}$$

Here there are two reactions and four species: enzyme E, substrate S, a bound combination of substrate and enzyme SE, and product P. There are also three complexes: S+E, SE, and P+E. As another example, consider a single-reaction CRN which represents autocatalytic self-replication:

$$F + X \rightarrow 2X \tag{4}$$

Here a replicator X reacts with some food molecule F to produce a copy of itself.

A few important distinctions should be noted regarding CRNs. First, reactions may be **reversible** or **irreversible**. A reversible reaction is indicated by a transition symbol $\rightleftharpoons$ as in Eq. (1), while an irreversible reaction is indicated by the symbol $\rightarrow$, as in Eq. (3) and (4). Reversible reactions are typically close to equilibrium and can occur in both directions, while irreversible reactions are those that effectively only proceed in one direction. Equilibrium reactions are always reversible and both directions balance each other, so there is no net conversion of reactants into products or vice versa.

It is also important to distinguish **elementary** and **non-elementary** reactions. An elementary reaction occurs in a single step, while a non-elementary reaction represents the net effect of a sequence of several elementary reactions, sometimes called a **reaction mechanism**. For example, the pair of enzymatic reactions in (2) and (3) may be elementary, but they can also be represented by a single non-elementary reaction $S + E \rightarrow P + E$, representing a two-step reaction mechanism. In fact, this representation of enzymatic catalysis is used ubiquitously in biology, where it is called the Michaelis-Menten scheme [313,314].

The dynamical effect of a CRN is to change the distribution of chemical species in time and space. The dynamics of a CRN are typically represented by differential equations that reflect the effect of **fluxes** across the different reactions. In general, reaction fluxes depend on the reaction kinetics, which in turn depend on particular details of the reaction volume, rate constants, external parameters, etc. Moreover, the dynamics of a CRN can be represented either at a deterministic level of mean concentrations (such dynamics are sometimes called a rate equation) or at the stochastic level of fluctuating counts of individual molecules (sometimes called a chemical master equation). The latter is particularly appropriate for small systems, in which fluctuations play a bigger role. The simplest dynamic rule, which is appropriate for a deterministic description of an elementary reaction in a well-mixed solution, is called **mass action kinetics**. The mass action expression for the flux $J$ across the forward reaction in Eq. (1) is

$$J = \kappa \prod_i [A_i]^{\nu_i} \qquad (5)$$

where $\kappa$ is some rate constant and $[A_i]$ indicates the concentration of species $A_i$ at a given point in time. For non-elementary reactions, so-called 'Michael-Menten kinetics' and other types of kinetics may be employed [313,314].

Finally, we briefly mention thermodynamic properties of CRNs. The most important thermodynamic quantity associated with a given chemical reaction is the **affinity** $A$, also called the (negative) **Gibbs energy of reaction** $-\Delta G$ [309], which indicates the increase of thermodynamic entropy due to the reaction. In accordance with the Second Law of Thermodynamics, $-\Delta G \geq 0$ for any reaction that occurs spontaneously (without additional energy input), and $-\Delta G = 0$ only for a reaction in equilibrium. For an *elementary* and reversible reaction, $-\Delta G$ obeys the following important flux-force relation,

$$-\Delta G = RT\ln(J^+/J^-) \quad (6)$$

where $J^+$ and $J^-$ are the fluxes across the forward and reverse side of a reaction, $R$ is the gas constant, and $T$ is the temperature. The flux-force relation plays a key role in chemical thermodynamics, since it relates reaction dynamics and thermodynamics.

The most interesting aspects of CRNs are often their non-equilibrium steady states, which are determined by the stoichiometric matrix as well as dynamical considerations. A common way to study such non-equilibrium states is using 'continuous stirred-tank reactors' (CSTRs). CSTRs are reaction vessels (simulated or real) in which a constant concentration of reagents is continuously added to the reactor while it is kept as a constant total volume. These conditions specify a non-equilibrium boundary condition, and ensure that all reagents and products are diluted out of the system in proportion to their concentration. These two conditions help simplify models of CRNs in such reactors, and they are easy to implement experimentally. These conditions have been used theoretically to model CRNs based on small organics (such as in the GARD model [315]) and evolutionary models [316] based on replicators, see Section 4.4.1. These conditions have been used experimentally to explore self-organisation in small organic reaction networks [317,318], and in macromolecular systems [319].

### 4.2.3 Chemical Kinetics Calculations

Chemical kinetics describes the rates at which reactions occur in a chemical system. In contrast to thermodynamic equilibrium calculations, which are time-independent and only describe equilibrium properties, chemical kinetics can be used to calculate the concentrations of various species in a chemical system as a function of time.

Consider the following reaction as an example:

$$a\text{A} + b\text{B} \rightarrow c\text{C} + d\text{D}, \quad (7)$$

where A and B are reactants with stoichiometric coefficients $a$ and $b$, respectively, and C and D are products with stoichiometric coefficients $c$ and $d$, respectively. We consider the general case of more reactants, products and reactions in Section 4.2.2.

The rate for the production of species C and D, and the rates of depletion of species A and B are then,

$$\frac{1}{c}\frac{d[\text{C}]}{dt} = \frac{1}{d}\frac{d[\text{D}]}{dt} = -\frac{1}{a}\frac{d[\text{A}]}{dt} = -\frac{1}{b}\frac{d[\text{B}]}{dt} = k[\text{A}]^a[\text{B}]^b, \quad (8)$$

where $k$ is the rate constant for the reaction in units that depend on the reaction order (e.g., s$^{-1}$, cm$^3$ s$^{-1}$, cm$^6$ s$^{-1}$), and the square brackets denote species concentration in units mol/cm$^3$. In this example, we have used simple mass-action kinetics, as appropriate for well-mixed ideal solutions, although other choices of kinetics are also possible [320,321].

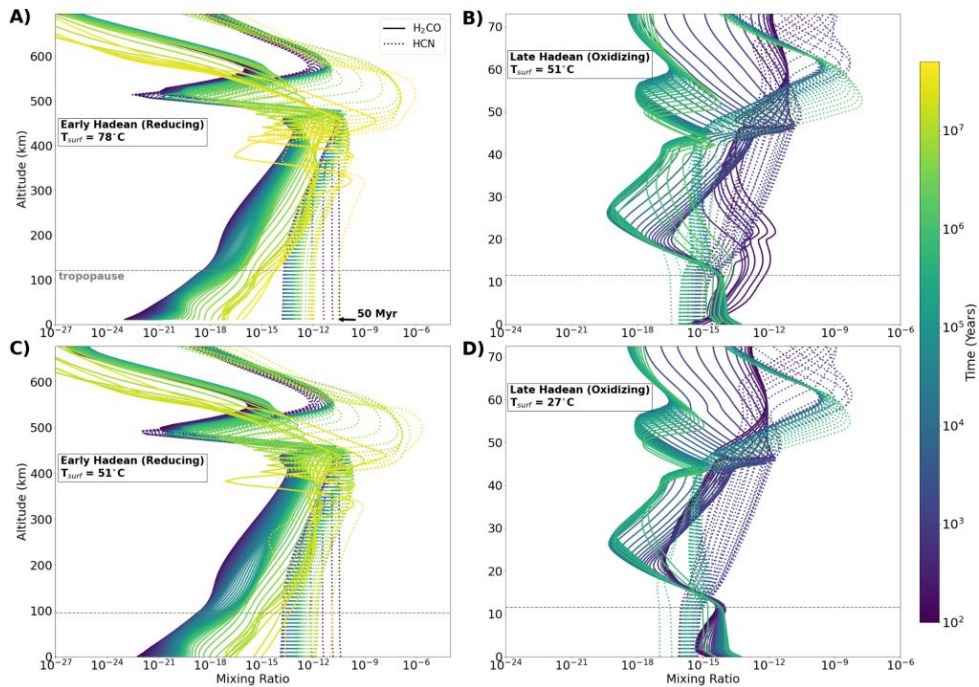

**Figure 4:** Four chemical kinetics models showing the evolution of hydrogen cyanide and formaldehyde in the Hadean Earth atmosphere. The Early Hadean model is at 4.4 billion years ago (bya), and the Late Hadean model is at 4.0 bya. Figure is adapted from [302].

It is more challenging to characterize the dynamics of an entire system than it is to characterize individual compounds or interactions. One way to approach system characterization is by constructing kinetic models that describe the system's behaviour and allow the modeler to make predictions. The first step is to use the current understanding of a system and its components to construct basic kinetic models that attempt to explain the observed behaviours of the system. Poor explanatory capacity of the model means that our current understanding of the system may be incomplete (or improved data is required). The next step is to improve and extend the model while seeking the simplest model that can successfully and reliably describe the data, according to the parsimonious principle (Occam's razor). It now becomes necessary to test if indeed the updated model can explain the data. Such a test can be done by performing data-fitting and parameter estimation, meaning to systematically explore alternative models and algorithmically testing their fit to the data [322, 323]. A recent example of such an approach comes from identifying a suprapolymerisation pathway of a self-replicating system [324].

A simple chemical kinetics simulation would involve numerically solving the system of ordinary differential equations above to obtain the concentrations of products C and D as a function of time. In practice, chemical kinetics simulations could solve networks of hundreds to thousands of chemical reaction rate equations simultaneously. In many cases, chemical concentrations in these simulations reach a steady-state solution after a certain amount of time; this is not to be confused with thermodynamic equilibrium, which has no reaction fluxes. Many open source chemical kinetics solvers exist, including Kintecus [325], Cantera [295], and ChemPy [326]. These solvers take as input a collection of chemical reactions with their temperature-dependent rate constants, as well as initial species concentrations, and the simulation time, and output chemical abundances under the assumption that the system is well mixed.

There exist chemical networks that contain collections of rate constants as a function of temperature calculated from experiments or quantum chemistry simulations or estimated using thermodynamics or similar reactions. Some examples are, CRAHCN-O [327–329], KIDA [330,331], STAND [332], and UMIST [333].

Chemical kinetics is also a common approach to calculate the chemistry occurring in planetary atmospheres in 1 spatial dimension (altitude). There are multiple codes that calculate the chemical kinetics and the physics (e.g., radiative transfer, diffusive mixing) for planetary atmospheres, which require additional inputs, including photochemical reactions with their dissociation cross-sections, the top-of-the-atmosphere solar spectrum, the altitude-dependent pressure and temperature profile, the altitude-dependent turbulent mixing profile, and influxes and outfluxes of species to/from the top of the atmosphere and planetary surface. A non-exhaustive list of chemical kinetics codes for planetary atmospheres can be found at [334] .

Chemical kinetics has been used for various OoL studies, including simulating HCN production in the atmospheres of early Earth [302,304,335–338], as well as organic synthesis in protoplanetary disks [339,340]. In Figure 4, we display chemical kinetics models of HCN and $H_2CO$ production in the Hadean Earth atmosphere from [302].

## 4.3 Graph/ Network Theory

In studying OoL processes it is often necessary to understand the interaction between individual components rather than just components themselves. In that effort, network models are indispensable. Inspired by graph theory in mathematics, network theory considers networks as sets of objects (nodes) connected by links (edges) [341]. Networks are often used in OoL studies to understand chemical systems, for example, by constructing chemical reaction networks from empirical or simulated data. The most common example of a chemical reaction network is one where there are two types of nodes: 1) chemical species (such as CO2, ATP, water, cysteine) and 2) chemical reactions which represent how sets of chemical species are converted to other sets of chemical species. In a network like this, a chemical species node would be connected to a reaction node if the species was involved in that reaction (see Section 4.2.2 for more on CRNs). Chemical reaction networks are mathematical objects each of which represents an entire set of reactions and compounds. By constructing these networks, we can analyse them to characterise the properties of the system and their interactions, rather than just the properties of the individual components. Walker and Mathis have reviewed how network theory can be leveraged to investigate the OoL problem [342].

### 4.3.1 Static Network Models

Once the rules for constructing a network are established and the network has been generated from the underlying data, its connectivity and statistical properties can be analysed [341]. Analyses based on the properties of a network's structure alone are often referred to as 'Static Network models' to distinguish them from properties that depend on some kind of dynamical laws imposed on the network (which we discuss below). Common statistical properties of interest in network analysis include the degree distribution (how many connections each node has), the average shortest path length (how far away nodes are from each other) and the average clustering coefficient (how densely connected nodes are). Another property of networks is its community organisation [343,344]. A community is a sub-part of a network which is more connected within itself than with other sub-parts, and is featured in many networks. Similarly, networks themselves, as wholes, can be characterised in several ways. In analysing static networks, it is crucial to develop appropriate controls. For example, the average shortest path of a network will only have meaning when compared with controls such as randomised versions of the network [342]. These are just examples of measures and properties that can be used to characterise networks, and the physical, chemical and biological implications of these measures will depend significantly on the specific system and hypothesis.

Constraint-based metabolic modelling is an example of static network generation that represents large metabolic systems at steady-state or dynamic equilibrium [345]. Such models have been used to infer essential and ancestral metabolic functions in prokaryotes, specifically tRNA-charging and cofactor metabolism [346]. Using these models, it is also possible to explore a combinatorial set of molecules and parameters associated with possible early Earth environments, and apply an enhanced network expansion algorithm to determine which proto-metabolic networks are thermodynamically reachable under different initial conditions. Such methods have given predictions about a possible phosphate-free metabolism [227] or an organo-sulphur nitrogen-free metabolism at the OoL [347].

### 4.3.2 Automated Reaction Network Generation

Automatic reaction network generation is a rule-based computational methodology. It reconstructs chemical reaction networks, from a limited set of predefined chemical transformations, called reaction rules, that happen on molecules due to reactions. Molecules are represented either symbolically, as in SMILES (Simplified Molecular-Input Line-Entry System) [348] or as molecular graphs (bond-electron matrices) [349]. Therefore, the reaction rules are defined either as symbolic or as matrix transformations. Reaction rules typically correspond to reaction types. For example, in the rTCA cycle, reductive carboxylation can be represented as one reaction rule capturing atomic transformations between acetyl-CoA and pyruvate, and between succinate and oxoglutarate [350].

Starting with a set of available molecules, reaction rules are algorithmically applied to the reactive sites of the molecules, consequently transforming reactants into products, which in turn can become reactants in the following iterations [351]. Through successive iterations, one expands the pool of available molecules and systematically stores their interactions as a reaction network. The feasibility of a reaction can be accessed by calculating the corresponding Gibbs free energy using either quantum chemical calculations [352] or group contribution methods [353,354], see Sections 4.1 and 4.2 for more details.

The automated reaction network generation methodology has multiple implementations. Examples of corresponding software are Netgen [355], RMG [356], Alchemy [218], and the latest pipeline from [357].

Following the work by [358], [350] used the automated reaction network generation to study the properties of the non-enzymatic rTCA cycle in the context of OoL and suggested that the rTCA hypothesis can be merged with the glyoxylate hypothesis [359–361]. Whereas, [218] used the automated reaction network generation to address emergent phenomena in prebiotic reaction networks and trace synthesis of life's building blocks from a set of chemical compounds that were present in the atmosphere of early earth.

### 4.3.3 Network Autocatalysis

In its traditional definition, autocatalysis refers to any reaction in which a product is also a catalyst, and such a product is called an autocatalyst [362]. Here, it is important to distinguish between an elementary reaction (e.g. A+X→2A) and a multistep reaction (e.g. A+X+B→A+C with B+C→A, whose combined result is A+X+2B→2A). Network autocatalysis is by definition a multistep autocatalytic reaction; however, the reaction steps can be very many and how reactions are wired can be highly complicated, such that a reaction network instead of a net reaction equation better represents such a system. Since all known forms of life convert external nutrients to more of themselves by a tremendous number of reactions (i.e., life catalyses the production of life), all life forms can be viewed as autocatalysts performing network autocatalysis. However, present-day life is not the only form of physicochemical systems that conduct network autocatalysis. Abiotic reaction systems, such as the formose reaction [363], the dissolution of copper in nitric acid [364], and the Belousov–Zhabotinsky reaction [365] are also examples of network autocatalysis, although their reaction networks are much simpler than a metabolic network.

As self-propagation is a shared feature between life and some much simpler abiotic systems, there has long been an interest in using theoretical models of network autocatalysis to explain the origins of attributes of life, to find candidate processes underlying abiogenesis, and to direct experimental studies on abiogenesis. These models include but are not limited to $(M, R)$ systems [366], hypercycles [367,368], the theory of reflexively food-generated (RAF) sets [228,369] that is based on the earlier conceptual model of collectively autocatalytic sets (CAS) [370], chemical organisation theory (COT) [371], systems of stoichiometrically autocatalytic motifs [214,372].

Another type of network-autocatalysis model is the GARD (graded autocatalysis replication domain), used to study life's origin in a Lipid World scenario [315,343,373–375]. The model demonstrates how life-like properties may emerge in a mutually-catalytic network of self-assembling amphiphiles forming assemblies such as micelles and vesicles which can collectively reproduce, store and propagate compositional information, and undergo adaptive evolution [375]. See [343] for a brief discussion regarding a criticism about the model's evolvability.

These models have different emphases and attributes. To help readers assess which models are suitable for their purposes, we briefly categorise these models in Table 4.3.1. Interested readers can also refer to [376] for a review of several models of autocatalytic sets, and to [377] for a broader discussion of the historical context centred on the concept of catalytic closure.

**Table 4.3.1. Categorising models of network autocatalysis.**

| Model | Catalytic polymers necessary? | Is catalysis of all nodes necessary? | Notes |
|-------|-------------------------------|--------------------------------------|-------|
| ($M$, $R$) systems | No | Yes | This model assumes that every component of metabolism has a finite life-time, so the system must be able to 'repair' the components by reactions. |
| Hypercycles (see also Section 4.4.1) | Yes | Yes | This model was originally proposed to explain how replication fidelity could be maintained without a long error-correction enzyme. |
| RAF (Reflexive Autocatalytic Food-generated networks) | No | Yes* (spontaneous catalyst can be added in Food to overcome this) | Every reaction needs to be explicitly catalysed by at least a chemical species in the network. Stoichiometry is not implemented explicitly. Based on Kauffman's CAS model. |
| COT (Chemical Organisation Theory) | No | No | COT's definition of self-maintenance does not guarantee autocatalysis, but it is possible to enforce autocatalysis by assuming environmental openness. Every reaction within a chemical organisation must have a positive flux. |
| Systems of stoichiometrically autocatalytic motifs | No | No | Not every reaction within an autocatalytic system must have a positive flux, so it is possible to find minimal stoichiometrically autocatalytic motifs by setting some fluxes to zero. Networks of autocatalytic motifs can be understood as ecological communities. |
| GARD | No | No | This model builds on catalytic networks and suggests |

| (Graded Autocatalysis Replication Domain) | | | replication and evolution of compositional information. |
| --- | --- | --- | --- |

Network-autocatalysis models have multiple applications in OoL research. For example, the theories of autocatalytic sets can help search for collectively propagating RNA systems [378]. One may also use RAF theory and COT to assess the conditions for a pathway to arise from simple nutrients to specific cofactors or more complex molecules [228,379]. In addition, minimal stoichiometrically autocatalytic motifs in databases of chemical reactions can be computationally detected, which makes it easier to design experiments aimed at constructing autocatalytic systems; and analyses of network autocatalysis could suggest possible routes of prebiotic evolution leading from simple systems to more complex ones [214,380].

## 4.4 Complex Systems Modelling

OoL research involves enormous chemical complexity, multiscale dynamics, and sometimes extreme environmental conditions, and therefore modelling is a valuable tool in its studies [381]. The most common use of modelling in the origin-of-life field is the exploration of the emergence of system-wide dynamic attributes, such as mutual catalysis, complexification and self-reproduction. Most chemical models employed in the field consist of a kinetic formalism ascribed to chemical networks, often in tandem with Monte Carlo and Gillespie stochastic algorithms, which are numerical methods that allow for the solution of kinetic systems of equations [382]. This can be seen in the study of collective autocatalytic sets (CAS)[315,370], replicator dynamics [383,384], and protocell models [385–387]. A growing body of work utilises differential equations, kinetic simulations, and agent-based models to explore nontrivial dynamics relevant to life's origin [343,388].

Other dynamical models are used to simulate entire cells, especially minimal cells developed from cellular genomes that do not include genes that are non-essential in the laboratory [389]. These models help us challenge what is minimally required for life and make predictions on the behaviour of such minimal cells.

### 4.4.1 Replicator models

In essence, prebiotic evolution faces the same problems as evolution of cellular life: information maintenance, storage, accumulation and its use to interact with other species. Because of this, prebiotic replicator models have a long history of influencing other fields and being influenced by them. Replicator models are usually based on the assumption that a molecular species (such as RNA) is capable of replication and mutation. Often these models abstract away the metabolic details of replication, and focus on its Darwinian evolutionary properties by providing an explicit selection criterion such as a fitness function, or an implicit evolutionary pressure, e.g. through cooperation.

One advantage of studying replicator models is that they decouple the origin of Darwinian evolution from the origin of cellular life, which enables the exploration of types of replicators or environments that have not been created experimentally, such as replicators in virtual geological environments representing mineral surfaces, ice crystals, and porous rock structures [390]. A growing body of research has constructed models displaying various characteristics of replicators and environments [383,384]. The approaches used to model replicators are many and diverse. In the following, we briefly describe some of the key models to show how replicator models could contribute to our understanding of prebiotic evolution.

Replicator models are typically formulated in terms of systems of coupled Ordinary Differential Equations (ODEs). Each equation represents the concentration of a molecular species, which changes according to its replication rate, through mutations, and through ecological interactions with other species. Important applications of this approach are the Quasispecies model and the Hypercycle model [368,391]. The evolutionary dynamics of replicator models are typically formulated in terms of mutation-selection mechanisms governing an evolving system which allows for much complexification. It consists of two parts: a matrix that specifies how each molecular species mutates into others, and a growth rate matrix that specifies how fast each molecular species grows in the absence of mutations. The mutation matrix — effectively a network that connects mutationally adjacent molecules — can describe the effect of realistic types of mutations, such as nucleotide substitutions, recombination, etc., while the growth rate matrix can be fitted to experimental data [392].

These models have shown that replicators with lower replication rate but higher connectivity in the mutational network can outcompete fast replicating but less connected species when mutation rates are high [393] (the so-called survival of the flattest effect). More recently, these models have clarified the effect of lethal mutations on the survival of the population [394], as well as the effect of selection acting on the phenotype of a species (for instance, its secondary structure) [395]. These models may be formulated in terms of differential equations, and are often integrated numerically, though some simplified cases admit analytical solutions [396]. Differential equation models treat populations of replicators as continuous quantities that represent mean concentrations, and are therefore not well suited for studying situations where small numbers of molecules are important.

Replicator models can be classified in terms of their growth kinetics [397–399]. In particular, the concentration of a growing replicator can be modelled with the differential equation $\frac{d}{dt} x(t) = c\, x(t)^p$ where $c$ is a constant and $p$ is called the 'order of a replicator'. The most common replicators are called 'first-order' replicators and have $p = 1$, resulting in exponential growth $x(t) = x(0)\, e^{ct}$. This growth is characteristic of autocatalytic reactions such as S + X → 2X, or networks of such replicators. Second-order replicators with $p = 2$ are characteristic of autocatalytic reactions S + 2X → 3X, or networks of such reactions, and lead to so-called hyperbolic growth. Second-order growth occurs when two replicators are necessary to replicate, as in sexual replication or other kinds of mutualistic interactions. On the other hand, so-called parabolic replicators have $p = 1/2$, such kinetics were originally observed in early experimental work on self-replicating templating RNA molecules [400,401].

It is important to emphasise that the order of autocatalytic growth is not just a mathematical parameter, but has important dynamical and evolutionary consequences. In particular, second-order replication leads to a nonlinear differential equation with much richer dynamics than first-order replication. For this reason, second-order autocatalysis plays a central role in several well-known models of bistability, oscillations, and pattern formation. This includes Eigen's hypercycle model of collective self-replication [368], the Schlögl model of bistability [402,403], as the core of the Brusselator model of oscillations [404], and the Gray-Scott model of pattern formation [405]. On the other hand, parabolic and other 'subexponential' replicators favour weaker selection with co-existence in steady state [406]. For a brief review of these issues, see [398,406].

### 4.4.2 Agent-based models

Individual-based models (also called agent-based models) describe populations of individual replicators, where each replicator is assigned some properties of interest that can determine its replication potential. This approach is closely related — but still distinct — to the replicator models. Upon replication, a copy of the selected individual is made, which can be mutated, thus allowing a Darwinian process to take place. These models can be extended more easily than their continuous counterparts to include more complex features.

One of the most successful extensions consists of introducing a genotype-to-phenotype (GP) map. In the case of RNA, fast minimum-free-energy folding algorithms from sequence to secondary structure [407] introduce one such GP map. This allows us to explicitly study RNA evolution as a process whereby mutations affect the genotype and selection acts on the phenotype (although there are conceptual ambiguities associated with genotypes and phenotypes embedded in the same molecule [408]). The evolutionary dynamics of populations of RNA replicators can be extremely rich, diverse and life-like - with complex patterns of mutational neutrality [409], evolvability and endless potential for innovation [367,409]. Encapsulating a population of RNAs into protocells, each containing a small number of RNAs, showed that it can withstand parasites much better than if there was no encapsulation [410]. Group selection at the protocell level can counterbalance template selfishness, as the protocells with a more balanced number of different RNAs (or with fewer parasites) can divide sooner, and even unfit protocells can stochastically re-generate a fit mixture of RNAs (i.e. stochastic-corrector- model).

In individual-based models that include spatial structure, replicators are assumed to occupy nodes on a lattice, representing a porous surface that limits diffusion. Discrete evolutionary rules representing the dynamics of the system in a simplified and computationally efficient manner. Replicators can increase in number by copying themselves into adjacent nodes in the lattice and interact with their immediate neighbourhood. Using such models, it was found that local interactions between replicators spatially limit the replication of parasitic templates [383,411] and generate emergent forms of organisation between cooperating replicators and their parasites [412,413]. While these models were originally developed as Stochastic Cellular Automata, they are now typically interpreted as spatially-extended agent-based models. This enables much complexification, for instance by equipping the replicators with a GP map as above [414,415]. Replicators can also be modelled to catalyse reactions and collectively generate a metabolism [416], and chemically variants of replicators, e.g. RNA vs. DNA, can be accounted for [417].

Alternatively, models based on evolutionary game theory have further simplified replicator dynamics to purely cooperative phenomena [418]. Game theory has been used to investigate micro-scale dynamics in the laboratory [419,420], and analogue lattice-based models have been used to show that cooperation could appear easily at the micro-scale [421].

### 4.4.3 Whole cell models

Conversely to Section 4.4.1, other approaches aim at modelling whole cells to predict phenotype from genotype and the environment using ordinary differential equations, providing a mechanistic understanding of how a cell works. As for genome-scale metabolic models, these models represent the function of each gene, gene product, and metabolite [389,422]. They also represent multiscale interactions at the cellular level; they consider the cellular structure, dynamic structure of molecular interactions, and the spatial compartment of the subcellular components [423]. Each subsystem is usually described with ordinary differential equations (e.g., kinetics), flux balance analysis, and stochastic simulations. To date, only two models have been developed: a model for the simple bacteria *Mycoplasma genitalium* [424], and a model for *Escherichia coli* [425]. More recently, a model of the 'minimal biological cell' JCVI-syn3A, a genetically minimal cell with only 493 genes and 452 genes coding for protein [426], was developed [389]. This represents the closest attempt at developing a near-complete whole cell kinetic model, 3-D spatially resolved, describing growth emerging from metabolism and gene expression (including the contribution of integral membrane proteins and lipids, etc.). This model couples ordinary differential equations and stochastic simulations to handle both the kinetics of metabolic networks and the kinetics of genetic processes, respectively. It describes the dynamic rates of genetic information processes, 148 known metabolites, 452 proteins and mRNAs, 29 tRNAs, 503 ribosomes, and DNA undergoing over 7000 reactions. The model gives quantitative insight into how the cell balances the demands of metabolism, genetic information, and growth over a cell cycle. The emergent behaviours arising from the simulations provide valuable understanding of the principles of life for minimal cells. Future development of similar models could be used in the context of the OoL.

## 4.5 Information-theoretic approaches

Information theory is the mathematical study of the quantification, storage, transfer, and use of communication (i.e. information). First formalised by Claude Shannon in 1948 [427], applications to OoL research have only garnered significant interest in the last few decades [428–430] . In what follows, we briefly review how investigating OoL-related questions using concepts related to information and, in particular, information theory, has contributed to shedding light on the transition from non-life to life, from the emergence of precursory mechanisms in physical and abiotic chemical systems, to new definitions of the living state.

### 4.5.1 Characterizing Evolution and Complexity

Evolution is the change in living systems across time, in informatic terms it is a processing of genetic information across large timescales. Kimura was the first to suggest that natural selection acts as an information acquisition process, and suggested a relationship between both information gain by natural selection and population growth by formalising the relationship between information and fitness [431]. Similarly, Schneider later investigated the ways in which living systems acquire information through evolutionary processes using Shannon information [432]. More recently, McGee et al. used ideas from computational learning theory and presented a revised and more robust version of Kimura's initial assessment [433], and Hledík et al. analysed information accumulation and maintenance in the genome using the Kullback-Leibler divergence—that measures the statistical distance between two distributions—noting its relevance for the OoL problem [434]. Concepts such as fitness and optimal information processing have also been studied along similar lines [435–437].

Further research on applying an information viewpoint in biological research has focused on protein, RNA and DNA. Examples include work done by Adami, who reviewed applications of information theory to protein evolution [438] and discussed applications such as predicting the secondary structure of ribozymes [439]. Assessing the complexity of sequences has also been investigated by Adami [440], and Yockey used Shannon's channel capacity theorem to analyse the problem of transmission in translated or decoded codons [441]. Similarly, information theory also fosters the study of how the genome has learned about the environment in the course of evolution; work completed by Jost has discussed how genetic information and external factors combine to create an organism [442], while informational properties that have allowed DNA to become predominant were examined by Travers [443].

In many regards life is seen as more complex than non-life. But in order for this observation to be scientifically useful, it needs to be paired with a concrete definition of complexity. Information theory has been a source of candidate complexity measures that have been used to characterise the differences between non-living and living systems. For example, information complexity can be a proxy for functional complexity [444], a view that is corroborated experimentally by Szostak's *in-vitro* work on ribozyme functionality [445]. Similarly, information-theoretic concepts have been utilised to investigate systems which display self-organisation [446], an important characteristic seen within living systems, and abiotic systems. Another avenue explored via information theory is the study of mechanisms within physical or chemical systems thought to underlie higher order biological functions such as behaviour. Larson et al. investigated such 'behavioural chemistry', arguing that molecules complex enough to attain a stochastic element in their structural conformation—such as RNA polymers of sufficient length—have the capacity to behave [447]. The various ways in which molecular organisation underlies the complex architecture of cellular life—such as molecular information processing—were likewise investigated by De La Fuente et al. [448].

### 4.5.2 From physical matter to biological matter

Information theory also allows quantitative analysis of the probability of emergence of primitive replicators from an abiotic environment. Since evolution fundamentally operates on information, we can calculate the likelihood of polymer assembly from random processes. Carrying such calculations on the rate of formation of monomers, Adami found that the likelihood could be extremely low. However, by also considering the biassed probability distribution in which they're found in functional informative molecules those probabilities increase dramatically [449]. These results are in agreement with earlier investigations by Yockey, who used the Shannon-McMillan-Breiman theorem to derive similar values [441]. While these issues are difficult to assess *in vivo*, experiments made *in silico* nonetheless have helped tackle these problems: using a model of replicators coupled to their environment, Mathis et al. showed that the transition from non-life to life could coincide with a phase transition measured through the mutual information—a metric that quantifies the mutual dependence between two random variables—shared between replicators and the environment [450].

Storing, communicating and predicting with information have been considered essential features of life [451, 444], and it is relevant to determing how these can emerge from the abiotic. To this effect, information-processing capabilities have also been shown to emerge as *non-neural learning* capabilities in physical and chemical systems. Investigating the Gray-Scott model of reaction-diffusion, Bartlett and Louapre demonstrated that chemical agents can spontaneously learn to react to their environment, thus implying a reduction on the required complexity for emergent learning [452]. Relatedly, Schnitter et al. have shown the capacity to store information in chemical reaction systems to be an emergent feature of the reaction cycle that forms crystals [453].

An early advocate for the role of information into abiogenesis is Küppers, who asserted that *all* life phenomena are steered by information—thus implying that the problem of the OoL is the origin of biological information [454]. Along the same lines, Davies suggested that to be successful, theories on the OoL must account for the informational capabilities of living cells [455], a view echoed by Nghe et al. who cites *information control* as one of six key parameters that should be investigated in prebiotic networks [418]. These questions have been explored numerically and conceptually in OoL literature. Probabilities for the 'emergence of information' have been calculated by Adami and LaBar from first principles by considering the minimum amount of information needed for self-replication [456]. Other scenarios for this emergence of information-processing capabilities include that of Rokhsar et al., who observed the spontaneous formation of biological-like phenomena in a spin-glass model [457], and Segré et al. who showed that small assemblies of spontaneously aggregating molecules could exhibit transfer of chemical information [458]. Another model was discussed by Kauffman and Lehman, who presented a scenario for the origin of biological coding where the latter would have originated from the cooperative relationship between collectively autocatalytic sets of nucleic acids and peptides [459].

Closely related fields to OoL research—that can in return inform research on abiogenesis—include work aiming at reproducing features of living systems *in silico* or *in vitro*. *In silico* experiments have so far been useful tools in investigating problems through an information-theoretic perspective, such as the accumulation of genomic information [460] or artificial cell design [461]. Other topics of investigation include information storage [462] and, in the realm of synthetic biology, advantages of multicellular consortia relative to information storage [463]. Of particular relevance for OoL studies, biological networks and organisation principles have also been subjected to investigation. For instance, Tasnim et al. has explored information transmission in complex networks typical of living systems [464], Walker et al. investigated the uses of information in deciphering hierarchical organisation in biological systems [465], and Kim et al. analysed Boolean models of biological networks to determine if information processing was an intrinsic property [466].

An informational perspective allows for an integration of the different fields of thought associated with OoL research by providing a common mathematical framework enabling cross-disciplinary collaboration. Such a perspective can thus be involved in every step leading from physical systems to biological ones by providing *quantifiable* measurements of complexity and information dynamics having been identified as a key property of living systems. A common framework is essential to answer such a multi-faceted question as the OoL problem.

## 4.6 Molecular phylogenetics

Molecular phylogenetics studies the hierarchical evolutionary relations, based on shared ancestry, between extant and extinct taxa. Darwin's 'Origin of Species" contains the first known phylogenetic tree representing the common origin of many taxa known at that time [467]. Molecular genetics and sequencing technologies (see Section 2.5) enable the construction of trees using the information embedded in DNA, RNA and protein polymeric sequences. Many applications in OoL research are common with other OoL problems that centre on the evolution of traits such as metabolic pathways and the cell physiology of extinct or unknown organisms. Research focuses in this area include the inference of LUCA characteristics based on phylogenetic analyses of genes [160], the transition from prebiotic world to biotic worlds [229,233], and the resolution of taxonomic relationships between bacteria and archaea [468]. Gene and genome trees have shed light on the origin and evolution of novel enzymes and novel metabolic pathways, and their ancestral states have been reconstructed by statistical evolutionary models [469,470]. Phylogenetic methods have been important to establish the ubiquitous nature of lateral gene transfer (LGT) between both ancient and extant organisms [471,472]. Other research has used phylogenetic methods to evaluate specific evolutionary hypotheses regarding the OoL. For example, previous work has demonstrated that the emergence of both LGT and the frozen genetic code likely predated LUCA, explaining how aminoacyl-tRNA synthetases were shared between branches of life [473,474]. Phylogenetic models have also been used to identify the root of the tree of life [475], helping to identify which extant organisms are most likely to share characteristics with LUCA. Similarly, phylogenomics has resolved origin of Eukaryotes within the Archaeal domain, leading to the conception of the two Archeal and Bacterial 'kingdoms' of life [476,477]. Finally, these approaches have reconstructed the evolutionary origin of specific metabolic pathways [478] and the possible metabolic capabilities of early life [160,479].

### 4.6.1 Homology and functional gene annotation

In phylogenetics, common ancestry is deduced from trait similarity. Yet, careful analysis is required to distinguish traits with common ancestry (homologous) from those with close function but different origin (homoplasy), or those sharing ancestry but being an evolutionary novelty (apomorphic). In molecular phylogenetics, homology in DNA, RNA and protein sequences is identified from sequence similarity. Homologous sequences that retain similar functions through speciation events are referred to as orthologs. On the other hand, genes that emerge via duplication events, referred to as paralogs, can acquire new or specialised functions in their subsequent, distinct evolutionary path. Paralogs can be an important source of information when building phylogenies and the distinction may be relevant when using phylogenetic trees for ancestral state reconstruction [480]. Also, sections of the genome can jump horizontally between evolutionarily distant taxa by LGT rather than by vertical inheritance. LGT can invalidate certain assumptions for tree building [481], distance matrix calculations, and, particularly, 'molecular clock' calibration [482,483]

Sequence homology underlies methods for gene and protein functional annotation in the absence of experimental characterization. There are several databases with sequences annotated for functions that are either experimentally known or imputed based on sequence information. Sequence matching algorithms [484], like OrthoMCL, inParanoid or Reciprocal Best Hits, are used to annotate genes or proteins of unknown function. Popular databases are Clusters of Ortholog Sequences (COGS), The Kyoto Encyclopaedia of Genes and Genomes (KEGG), EggNOG and OrthoFinder among many others[226,655,656]. Popular servers for prokaryotic gene annotation are RAST or MEGAN [485,486]. Similar methods are used to annotate protein function in Metagenomic Assembled Genomes (MAGs) in environmental metagenomics studies [487]. Other methodologies permit the reconstruction of metabolic networks from functional annotation by using orthologs linked to metabolic pathways, such as KEGG mapper [488], COG pathways [489] or modelSEED [490] and RAST server to search feasible metabolisms in a given environment [490–492]. Ortholog detection algorithms also help to identify LGT in microbial evolution [493,494] and date evolutionary events [495,496].

### 4.6.2 Constructing Trees

Phylogenies are built by comparing traits among taxa. Traits are typically morphological, either coded as binary, e.g., presence/absence of an organelle, or numeric, e.g., number of cilia, or functional, e.g., ability to produce/consume a given metabolite or certain amino acid metabolic pathway synthesis [497]. In the case of sequence-based phylogenies, the corresponding traits would be mutations in the DNA, RNA or protein sequence, i.e, nucleotide/amino acid substitutions or insertions/deletions. Sets of informative traits are collected in tables and passed to phylogenetic tree building algorithms that use either the original character matrix or a derived distance matrix. Distance matrices are double entry tables filled with values proportional to the number of sequence changes calculated after sequence alignment. In pairwise sequence alignment, all combinations of sequence pairs are arranged so their homologous sites are adjacent, either using global [498] or local alignments [499]. Molecular phylogenetic reconstruction uses multiple sequence alignments (MSA) constructed by algorithms like CLUSTAL-Omega [500], MAFFT [501], and MUSCLE [502]. There are recent alignment-free methods usable to calculate distance matrices from molecular data [503,504]. Finally, some alignment methods leverage the greater conservation of protein structure over sequence [505] and integrate available structural information into the alignment procedure [506].

A second task required for modern phylogenetic inference methods is the selection of a best-fit molecular evolutionary model. These models integrate features of molecular evolution, including the rates of pairwise nucleotide or amino acid substitutions, evolutionary rate heterogeneity across different positions in the MSA, and base frequencies of each nucleotide or amino acid. Model testing can be carried out by maximum-likelihood-based methods which calculate the likelihood (i.e., probability of the extant sequence data given the model) of different candidate models. Model selection criteria consider the likelihood of candidate models while penalising overparameterization. Software for model testing include ModelTest-NG [507] and ModelFinder [508].

Given both an MSA and best-fit evolutionary model, phylogenetic reconstruction tools infer a tree by resolving branching patterns and branch lengths. Like model testing, tree searches can also be performed by maximum-likelihood-based methods which calculate the likelihood of candidate trees during the search process (e.g., RAxML [509], IQ-TREE, [510]). For nearly all real-world sequence datasets, possible tree space is massive and computationally infeasible to explore completely. Therefore, heuristic approaches are used to identify the most likely tree. Branch support can be assessed by different metrics, including nonparametric bootstrap (evaluating the frequency of given clade, or cluster of sequences, across different trees reconstructed from resampled sequence data) [511] and likelihood ratio tests, which compare the likelihood of the best tree with that of the next-best tree with the branch in question collapsed [512]. Finally, phylogenetic reconstruction can also be performed through Bayesian inference. Bayesian tools empirically approximate the posterior probability distribution across model parameter space (including both the tree and evolutionary model) by sampling parameter values through Markov chain Monte Carlo-based (MCMC) algorithms. These methods thereby identify the maximum a posteriori tree as well as provide a measure of tree uncertainty. Branch support is expressed as the posterior probability of a given clade, or the frequency of the clade across sampled trees. We direct the reader to Holder & Lewis [513] for a discussion of different phylogenetic inference methods.

### 4.6.3 Molecular clocks

Emile Zuckerkandl and Linus Pauling first suggested the 'molecular clock' hypothesis: if the mutation rate in nucleotide and protein sequences correlates linearly with time and remains relatively constant under neutral evolution (Kimura (1987), this relationship can be used to date evolutionary divergences [514]. The original 'strict molecular clock' method measures lineage- or sequence-specific parameters (e.g. the substitution rate per year). This can be computed as, for example, a linear model between genetic distances and time divergence [515]. Later models control for branch specific substitution rate (i.e. 'multi-rate' and 'relaxed' clocks), and advanced versions estimate the parameters by means of MCMC Bayesian parametric statistics, maximum-likelihood methods and others [516]. When possible, models are time-calibrated by including nodes dated from fossil records [517]. For example, fossil lipids can be used to calibrate molecular clocks early in the evolution of microbes [524]. In some cases, calibrations can be based on geological or climate data [518], and, for rapidly evolving taxa with poor fossil records, calibrations can be based on estimated molecular substitution rates or sampling dates [519]. Recently, it has been shown that LGT can be used to date phylogenetic events such as bacterial radiation [495] and methanogen evolution [520]. Molecular clock models and phylogenomics can help date major events in the evolution of microbial taxa, including the origin of Archaea [521] and LUCA [522,523].

### 4.6.4 Ancestral reconstruction

Ancestral sequence reconstruction (ASR) was introduced by Pauling and Zuckerkandl [525] as an application for molecular phylogenetics, molecular clocks and orthology relations [526]. ASR is a method for inferring the sequence content of ancestral proteins or genes corresponding corresponding to internal nodes of a phylogenetic tree. Typically, a researcher is interested not only in the ancestral sequence itself, but of its phenotypic outcome (e.g., the biochemical or biophysical properties of a particular ancestral protein). With modern gene synthesis services, it is practical to synthesise the encoding gene of an ancestral protein and 'resurrect' it in the laboratory by expression in a modern host organism, followed by phenotypic characterization [527,528].

ASR relies on many of the same computational methods that underlie phylogenetic inference. Earliest ASR studies relied on maximum parsimony methods [529,530], which seek to minimise the number of substitutions in a phylogeny [531]. ASR is more commonly performed today by probabilistic methods like maximum-likelihood (e.g., PAML [532]) or Bayesian inference (e.g., MrBayes [533]).

Regardless of the method, uncertainty in ASR is a key consideration for downstream analysis. Even for relatively small genes or proteins, the probability of the reconstructed ancestral sequence will typically be very low. For a 100-amino-acid protein, even if the reconstructed residue at every site has a probability of .9, the probability of the entire sequence (the joint probability of all residues in that sequence) = $.9^{100} \approx$ 3e-5. Thus, much work has been focused on understanding the significant sources of ASR inaccuracy, including uncertainty associated with multiple sequence alignment, evolutionary model parameters, and phylogenetic tree topology [534–536]. Nevertheless, previous experiments have demonstrated that phenotypic properties of interest can be robust to ancestral sequence uncertainty. Therefore, in these cases, the reconstructed sequence need not be accurate to draw scientific conclusions at the level of phenotype. A useful strategy is to integrate phenotypic characterization across a range of plausible ancestral sequences or to incorporate a 'worst-case' alternate ancestor, constructed by replacing any ambiguously reconstructed residues with their second most probable residue [535,537].

Because ancient proteins do not, except in relatively recent and rare cases, leave direct fossil records, ASR is a powerful tool to infer the properties of early-evolved proteins and metabolic processes that have been central to the development of the biosphere. For example, ASR has been used in both *in silico* and experimental studies to investigate the temperature stabilities of ancient proteins dating as far back as LUCA [538,539], the specificity and functionality of early translation machinery [540,541], the development of the genetic code [542,543], and the evolution of key metabolic processes and biogeochemical cycles [544–546].

For OoL studies, ancestral reconstruction also need not only be applied to molecular sequence information. Similar statistical approaches can be leveraged to directly reconstruct ancestral phenotypic traits, given a matrix of trait information associated with extant proteins or taxa rather than a multiple sequence alignment. This approach, referred to more broadly as ancestral 'state' reconstruction, has been used to infer the minimal gene-set of LUCA [547], the cell shape and taxonomic affinity of the last bacterial common ancestor (LBCA) [233], and the ecological attributes of earliest photosynthetic organisms [548,549]. Future work in this area may incorporate recently developed methods for inferring features of the complex cellular networks early in life's history, including protein interactions [550].

# 5. Bridging theoretical and experimental approaches

As the questions related to the origin(s) of life (OoL) have evolved, they have incorporated more heterogeneous sources of data and more sophisticated techniques. For example, the rise of Omics approaches enabled the systematic investigation of biochemistry, led to testable laboratory hypotheses, and has facilitated new approaches to systems chemistry. Similarly advances *in invitro* selection experiments pioneered in evolutionary biology are being applied to chemical systems to understand selection in proto-biological systems. New technologies have provided OoL scientists with new tools, leading to new approaches, as is the case with automation of laboratory experiments. A common trend in these tools is the feedback or interaction between experimental workflows and large datasets or modelling approaches which provides the opportunity to test old hypotheses while continuing to nurture new theoretical models (Figure 5). We consider these types of tools, which allow for tighter integration of computational workflows, and experimental approaches to be particularly important for the future of the OoL field. Therefore, we have dedicated a final section to address the comprehensive range of concepts and methodologies that, while not techniques themselves, act as conceptual bridges (Figure 5), enhancing the understanding and simultaneous application of different knowledge from various fields.

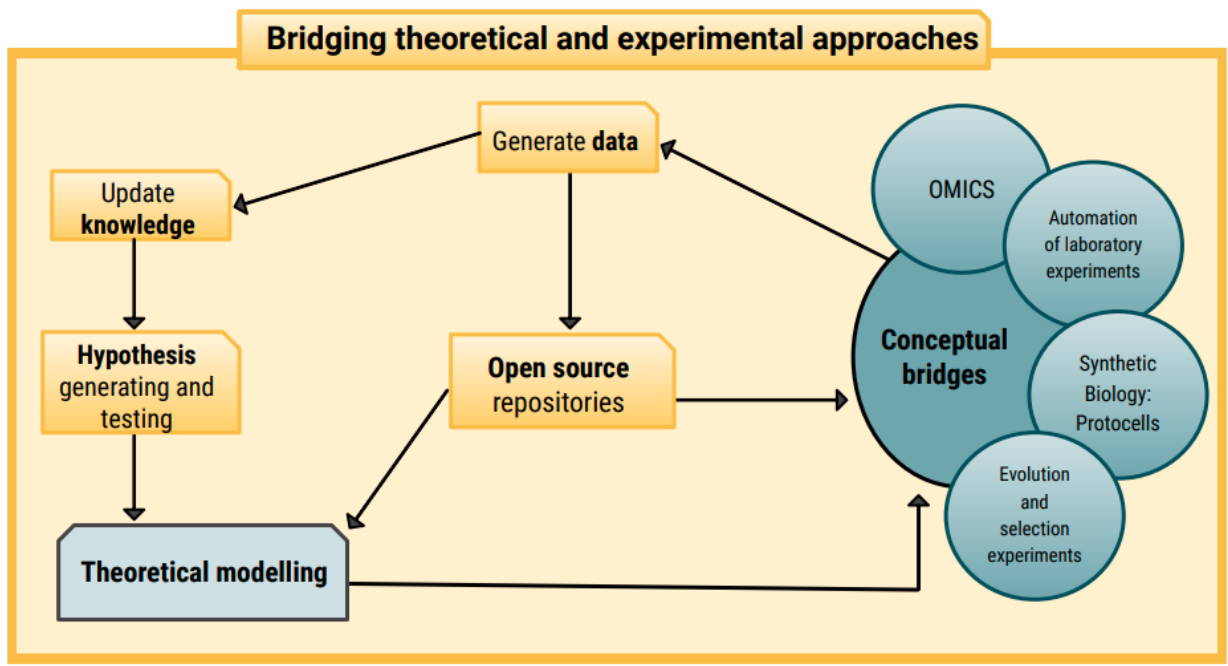

**Figure 5. Bridging theoretical and experimental approaches:** interplay between experimental workflows, modelling approaches and large datasets (conceptual bridges) enables the testing of hypotheses while fostering the development of new theoretical models.

## 5.1 Omics

Omics refers to a variety of biological disciplines (metagenomics, metabolomics, proteomics, transcriptomics, etc) whose overall objective is to describe and quantify cellular biological molecules indicating their composition, structure, dynamics, and function [551]. The emergence of -omics techniques has impacted many questions and fields in biology, including OoL, for instance in the study of LUCA [8,552] The diversification of omics approaches enables the exploration and understanding of different perspectives within various fields: Metagenomics focuses on nucleotide sequences, Proteomics and Transcriptomics analyse proteins and RNA sequences, while Metabolomics investigates metabolites [553–556].

### 5.1.1 Metagenomics

One of the most widely used techniques to study the structure and function of nucleotide sequences in environmental samples is metagenomics [557]. It generally consists of analysing heterogeneous microbial communities to obtain their genomic composition (i.e. both in terms of taxonomic and functional composition). Functional metagenomics [558,559] studies involve cloning these previously identified DNA fragments from the environment, expressing them as genes in a model organism, and screening for enzymatic activity. Metagenomics is particularly relevant when studying extremophile communities to identify genes with important and shared functions that are present within the community and to determine their adaptive pathways in the face of extreme conditions, which may be similar to those in early life [560]. Metagenomic analyses are especially useful due to the fact that these microorganisms do not have to be cultivated under laboratory conditions [561], which is critical since the extreme conditions in which these microorganisms are found are almost impossible to replicate in the laboratory [562,563]. For instance, the optimum salinity range for cultivable and isolated microorganisms is between 0 and 35% [564].

### 5.1.2 Proteomics and Transcriptomics

Proteins likely played a crucial role during the origin(s) of life on Earth [565]. There are two key -omics fields related to proteins: proteomics and transcriptomics. Proteomics refers to the study of the structure and function of the complete set of proteins expressed in an organism, also known as the proteome [566]. On the other hand, transcriptomics focuses on the comprehensive set of RNA molecules, both coding and non-coding. Over the past decade, the evolution of various techniques has contributed to the advancement of proteomics. The general principle involves characterizing the proteins encoded by a genome and their expression [567,568]. Mass spectrometry is the most widely used technique in this field (described in Section 2.2), which enables the high-throughput profiling of proteins through electrophoresis or isotopic labeling [569].

Transcriptomics aims to quantitatively assign reads to transcripts within a given genome. Early work in transcriptomics involved the use of DNA microarrays [570]. However, the most common contemporary technique for characterizing the transcriptome is RNA sequencing (RNA-seq). Its greatest advantage over previous methods is the ability to detect both known and unknown genes without the need for prior knowledge [571]. Single-cell transcriptomics has further enabled Single-cell RNA sequencing (scRNA-seq), allowing for enhanced analysis at the level of individual cells. For more information on techniques, we recommend reviewing [572] on transcriptomics.

These techniques have proven to be invaluable in the study of OoL, particularly in relation to the RNA world or protein-first hypotheses [565,573]. Proteomics and transcriptomics studies provide deeper insights into early life and evolution than genetic analysis alone, allowing for the investigation of fundamental components of the cell, such as the reconstruction of eukaryotic chromatin evolution [574] or the evolution of histones [575,576]. Furthermore, the study of structural domains of proteins holds significant potential in OoL research, such as in understanding the early origins of viruses [577].

### 5.1.3 Metabolomics

The identification of metabolites, which are low molecular weight organic compounds produced by living cells (both in pure cultures and complex microbial communities), is of significant interest in OoL studies, particularly when considering the complexity of prokaryotic and eukaryotic biology. The metabolome of a microbial community is not only influenced by its genetic capacity, which can be studied using sequencing techniques (see Section 2.5), but also varies based on environmental conditions such as temperature, pressure, pH, salt diversity and concentration, and irradiation exposure [578–582]. Metabolomics studies focus on various research objectives: i) investigating metabolites in environmental samples found in meteorites, stromatolites, or fossilized matrices on a smaller scale; ii) studying metabolomic pathways of existing microorganisms with primitive traits and theoretical pathways of potential primary metabolomes to understand early Earth conditions (e.g., without O2, with sulfur intake, in hot environments) on a medium scale; and iii) comprehending and reconstructing the complete metabolome of microorganisms within a population and their interaction with the surrounding community on the largest scale. In fact, several hypotheses about primitive cells and contemporary extremophiles arise from the interactions (e.g., symbiosis) between individuals [85,583–585].

Metabolomic studies employ a wide range of analytical techniques to detect and identify biochemical compounds (see Sections 2.1 and 2.2) [586,587]. These techniques include combinations of analytical methods such as i) High-Performance Liquid Chromatography linked to a Mass Spectrometer (HPLC-MS) [76,588], ii) Gas Chromatography linked to a Mass Spectrometer (e.g., GC-MS) [589,590], iii) Mass Spectrometer–Mass Spectrometer (MS/MS) [81,591], and iv) Nuclear Magnetic Resonance (NMR). Sample analyses can be enhanced with extraction methods, isotopic labeling and screening, capillary electrophoresis, and microfluidics to improve the detection and identification of various organic compounds, ranging from light to complex compounds [592]. These sample pre-treatments save time, aid in the better separation of organic or enantiomeric compounds, and enable the extraction of a wide and quantitative range of organics from large to small sample quantities, which subsequently become detectable [593].

Among the techniques, High-Performance Liquid Chromatography coupled with Mass Spectrometry (HPLC-MS) is the most commonly used and convenient method, allowing for the analysis of polar and non-polar or organic and aqueous phases independently or consecutively. Moreover, the high sensitivity of mass spectrometry provides a high level of certainty in detection (ranging from nmol to pmol) and enables additional isotopic analyses on the same dataset. These studies, in combination with complementary microscopy and statistical analyses, help establish connections between molecular compounds and biological activities influenced by the community, abiotic stress factors, and the genome. MS-based metabolomics is used to complement metabolic pathways that may not be detected through genome annotations [212], demonstrate the function of theoretical pathways and metabolites in cells or biofilms [212,594], discover novel metabolic pathways [595,596], and describe new biomarkers [597,598].

## 5.2 Automation of laboratory experiments

Automation can help overcome experimental limitations and can generate vast amounts of data. Most laboratories have a range of tools that can automate specific tasks, usually procedures of sequential repeating actions. Automation is employed more often in biological research than in chemistry, though its most common usage is for molecular procedures such as real-time PCR experiments and next-generation sequencing (Section 2.5.2) [599]. The advent of liquid-handling robots enabled efficient automation of more complex protocols [600,601], from peptide synthesis to enzyme-linked immunosorbent assays, and often utilising microfluidics capacities (Section 5.3) [602]. Population-level biological methodologies are sometimes automated as well, such as in culturing of microorganisms, community control and high-throughput genetic modifications [603-606]. Arguably, automation technologies are even more crucial for synthetic biology (Section 2.3), where reliable models require large amounts of high-quality data that can only be generated by automation, thoroughly testing a large diversity of systems with high reproducibility [607]. Lastly, while much of the detailed automated procedures are applicable to chemistry research, automation of complete chemistry experiments are somewhat hard to find. A few groups have started developing more advanced automated projects, but those are focused on drug discovery [608] or organic synthesis [609].

While automation can overcome human limitations and solve problems such as reproducibility [610], it is not widely used by researchers for exploratory experiments in the OoL field. One of the first examples in the field was an automated platform for peptide coupling [611] in which the synthesis of glycine oligopeptides in high yield was demonstrated. The same laboratory developed a microfluidic platform for the study of osmotically driven droplet growth and population behaviour [612]. Another group reported a system, which allowed the performance of automatically controlled wet-dry cycles [613] under anaerobic conditions. A completely different example is the development of a 3D printed stirring device, which further leverages the already established automated functions of an autosampler and enables dynamic combinatorial library experiments with direct injection into the analytical system [614]. While all those examples are steps towards more automated OoL experiments, they still require a high amount of human intervention within the experimental process.

In a new approach, a network-driven exploration was combined with an automated experimental platform with a closed-loop analytical system [615]. In this workflow, cyclic reactions have been performed over several weeks and decisions throughout the experiment have been taken by an algorithm based on real time analytical feedback. For researchers interested in the automation of laboratory processes, useful resources discussing the development process are available [616].

## 5.3 Synthetic Biology: Protocells

One way to understand how non-living molecules could turn into life is to reconstitute a living cell from scratch, a central work of bottom-up synthetic biology. Aside from being a topic of independent research [617,618], protocells have emerged as invaluable experimental and theoretical tools for exploring the fascinating stage between non-living matter and cellular life. Classically, biological molecules, such as nucleic acids, peptides, and proteins, can be encapsulated in synthetic compartments to build a model protocell that shows life-like properties [619, 620]. While using defined (purified) molecules, it is possible to describe in detail how dynamical properties emerge without many unknown factors seen in complex extant cells. The simplicity of protocells also allows a direct comparison with computational and theoretical studies [621]. Notably, biomolecules have been studied individually and in chemical networks possibly more than any other type of complex chemistry, and they are the only type of chemistry directly associated with life. As such, synthetic biology can help us find a minimal requirement for the emergence of life-like behaviours.

There are several common methods to generate protocellular models, especially lipid vesicles and liquid–liquid phase separated (LLPS) droplets [622-624], among others [390, 625]. Their formation often relies on the self-assembly of lipids or biopolymers, associated with the encapsulation of desired molecules. In the case of phospholipid-based vesicles, the simplest preparation method is to add an aqueous solution on a dried lipid film on a glass surface for swelling the film, leading to spontaneous vesicle formation, although the controllability of vesicle structures and the efficiency of macromolecular encapsulation are limited. Another method is to transfer water-in-oil droplets (with a phospholipid monolayer) through an oil-water interface (another phospholipid monolayer) to form vesicles with phospholipid bilayers. Vesicle structures and molecular encapsulation can be well controlled during the preparation of droplets. Microfluidics (Section 2.3) is also used to generate vesicles with extremely controlled sizes. Protocell structures and molecules to encapsulate are chosen depending on what phenomena one would like to see, and at what level of complexity. For example, combining short RNA with fatty acid vesicles could allow the exploration of possible ancient cellular self-reproduction [626], whereas integration of an artificial genomes and proteins with phospholipid vesicles, allows us to probe more complex cellular behaviours, which LUCA may have displayed [620]. In the latter case, most central biological functions, such as genome replication, protein translation, metabolism, and energy generation have already been reconstituted, at least to some extent. Non-biological cell-like compartments (e.g., water-in-oil emulsion) are also used to investigate the role of cellular structures in a more conceptual fashion, such as in molecular evolution (cf. Section 5.4) [627,628]. Although most studies in this field focus on membranous compartments, recent progress on reconstituting key biological reactions, including ribozyme catalysis, genome replication, and protein translation in LLPS droplets [629, 630], also highlight the versatility of membrane-free compartments as protocell structures [98].

## 5.4 Evolution and selection experiments

Evolution generally requires the following three steps. (1) Replication: copying of a molecule to another which inherits its trait. (2) Mutation: imperfect replication leading to diversification of traits, caused by replication errors or environmental perturbation. (3) Selection: a replicator with a particularly advantageous trait propagates more than others in a population. Similar to any living system, previously constructed molecular systems that can undergo these processes generally have replicable nucleic acids [631], such as RNA, as a carrier of genetic information that determines their traits. In this case, the mutation is equivalent to the change of nucleic acid compositions.

Once a molecular system that replicates and mutates is available, one might witness natural selection and evolution. A typical evolution experiment is performed through serial dilution or under continuous flow, with the supplement of nutrients (substances required for replication). During the experiment, repetitive replication generates mutated offspring with different traits, and dilution essentially causes the death of some and makes room for the progenies to reproduce within finite resources and space. If a replicator that can replicate faster emerged, it would dominate the population through successive replication and dilution, i.e., evolution occurs. The first *in vitro* evolution was reported in 1967 by Spiegelman's group, using an RNA with the supplement of a purified RNA polymerase to facilitate replication, and NTP [632]. The RNA gradually improved its replication efficiency in the course of evolution, while becoming shorter and shorter. Various evolutionary phenomena, including diversification, complexification, and niche partitioning, have been observed for more complex replicators, such as an RNA that replicates using its self-encoded RNA polymerase [633] and a catalytic RNA (ribozyme) that replicates using reverse-transcription (RT) and transcription enzymes [634]. A system may need to be encapsulated in a compartment to link the encoded information with its replication, akin to a contemporary cell. A similar experiment is also feasible without a protein enzyme. For example, a ribozyme that makes a copy of it by ligating RNA fragments can essentially evolve if pre-mutated fragments are supplied [635]. Although its evolutionary potential is not high due to the limited variety of mutations, it will increase if an RNA polymerase ribozyme capable of sustained replication is engineered in the future. Using synthetic chemical replicators, it has been demonstrated that replicators can spontaneously become more complex under out-of-equilibrium conditions [636]

In addition to the Darwinian evolution experiments described above, the evolutionary principle can be applied to screen DNA, RNA, peptides, and protein molecules with desired functions. Some techniques use biological organisms such as bacteria and viruses, whereas others complete the entire experiment *in vitro* [637-639]. The latter, known as *in vitro* selection, has been used to identify functional RNA and peptides possibly relevant to the OoL because of its ability to screen a large number of molecules and compatibility with non-canonical environments. A typical *in vitro* selection experiment first prepares a pool of randomised DNA sequences with constant regions for PCR amplification, followed by *in vitro* transcription to synthesise RNA depending on what to select. For screening nucleic acids, they are subjected to a certain selection (e.g, ligation or binding with a substrate), and 'selected' molecules are collected (e.g., by collecting the substrate attached to the molecules of interest). Compartmalising each molecule also facilitates the selection of complex activities, such as trans-reaction and multiple turnovers [640]. For peptides and proteins, each molecule is first cell-free translated from an RNA library in a compartment or via attachment to RNA that encodes each protein, usually through conjugation with a linker (mRNA display) or without releasing from the ribosome (ribosome display). These processes ensure the linkage of protein activity (phenotype) with its genotype (RNA sequence), not necessarily applying to the selection of nucleic acids as they can possess catalytic activities by themselves. After selection, sequences are amplified by RT-PCR to obtain a new pool. Through multiple cycles of these procedures, molecules with target functions gradually enriched from even a single copy in the population, similar to evolutionary processes. At any point, a pool can be sequenced as described in Section 2.5 and subjected to biochemical characterization. For OoL research, these techniques have been used to find ribozymes that catalyse biological reactions, including ligation [641], polymerization [642], and aminoacylation [643], and recently peptides, such as an RNA-binding peptide of prebiotic amino acid composition [644].

Evolution and selection experiments can also be performed on living organisms. These experimental evolution approaches often include extreme conditions as a selection source and model organisms such as bacteria: *Escherichia coli* [645] or yeast: *Saccharomyces cerevisiae* [646] among others. The analysis of the adaptive pathways to extreme conditions and the great genomic knowledge that exists about these model species facilitate the study of the OoL and evolution of the genome [647], for instance, DNA repeats [648], or genomic plasticity [649] in prokaryotes. Similarly, new computational and *in silico* approaches are combined with *in vivo* experiments to study the origin of adaptations [650].

## 6. Future directions and conclusion

Developing a scientific explanation for the spontaneous animation of matter from inanimate matter, and experimental demonstrations of that explanation will represent a landmark achievement in the history of science. If this is accomplished in the coming decades it will depend (at least in part) on the experimental and theoretical methodologies described here. But it is unlikely that an achievement of this scale will be accomplished within the confines of a single academic lab, or research group – solving the Origin(s) of Life (OoL) will require a high level of coordination and collaboration amongst distinct teams. Massive global collaborations have been organised in other fields such as particle physics, cosmology, and astronomy - but they're rare in the fields closest to OoL, such as chemistry and molecular biology. Establishing such collaborations will require a significant reorganisation in the way OoL science is funded, which is likely beyond the scope of individual researchers (particularly at the early career stages). But a necessary precondition for the success of such collaborations is the establishment of community standards for exchange and validation of data and software.

The OoL community must identify and adopt rigorous standards for sharing and distributing scientific data. This is an area where OoL science should follow the lead of different scientific communities by adopting standards which are already working. For example, in many fields, such as geosciences, the FAIR (Findable, Accessible, Interoperable, and Reusable) data standard is already becoming the norm [651,652]. The principles of FAIR data are general enough to be applicable to all fields in OoL;Done the challenge for early career scientists now is to implement these principles in a way that is compatible with their own disciplinary background and the broader OoL community. Similarly software and scripting are critical for producing and understanding scientific results, particularly as data sources and analysis pipelines become more sophisticated and complex. To properly document and distribute analysis code the OoL community should look to other disciplines for resources, such as the Turing Project which started in Data Science but is generally applicable to quantitative sciences [653]. Finally, as experimental procedures become more sophisticated, and as their results produce more life-like phenomena it will be essential that those results are clearly documented and reproducible. The development of universal standards for experimental procedures may be an effective tool to document and share these processes [654]. The OoL community should decide which of these tools and standards are useful for our science, but we should not continue a process of hidden data, undocumented code, and unreproducible procedures. The tools we need to improve the rigour and quality of our science exist, now we have to adopt them.

If scientists working on OoL hope to transition the field from one dominated by isolated (and dogmatic) research paradigms to a productive scientific community they must be able to communicate their advances with each other effectively. We hope that this review provides a starting point for those technical discussions. By focusing on empirical and theoretical results, within the context of their technical validity, rather than in the context of conceptual narratives, we hope to accelerate discovery and conceptual synthesis in the OoL community. While this review cannot cover each technical topic in the full depth and detail required, we hope that it serves as a roadmap for future OoL scientists to begin their journey of understanding the diverse topics in the field.

# Acknowledgements

This work is a collaborative effort of the titled authors as part of the Origin of Life Early Career Network (OoLEN. https://www.oolen.org). We chose to add OoLEN as the first author to give a better representation of this team effort, rather than listing any single author as the first author. We hope such a thing can be adopted by others. We indicate that authors 2-8 (AS, BC, BD, CRA, MC, MO, and PZ) have made a more distinct contribution. All authors are listed alphabetically by their last name. We would like to acknowledge all current and past members of OoLEN for their contributions to our community and to this manuscript. In particular we would like to acknowledge Evrim Fer who helped with Section 4.5.

# Author Contributions and Funding Information

- (AS) Silke Asche: Wrote Section 5.2 and parts of 2.1, 2.2 and 5. Contributed to figure 1 and made figure 2. Significant comments and edits.

- (AV) Avinash Vicholous Dass: Contributed to 2.2, specifically Subsection 2.2.1 Molecular Analysis. Section 2.1.3, Section 2.6, Subsections 2.6.1 X-ray diffraction and 2.6.2 Raman Spectroscopy

- (BA) Arnaud Buch: Reviewed Sections 2.2 and 5.1.3

- (BC) Carla Bautista: Contributed one figure in Section 5 (Bridging theoretical and experimental approaches). Original draft for Section 5.1 (Omics), Section 5.1.1 (Metagenomics), and Section 5.1.2 (Proteomics and transcriptomics). Review and original contributions on Section 5 (Introduction to Bridging theoretical and experimental approaches), Section 5.1.3 (Metabolomics) and Section 5.6 (Evolution and selection experiments). Review and small edits on Section 2.4 (Genomic Sequencing), Section 3.2. (Biochemical and Biological databases) and Section 5.4 (Automation of laboratory experiments). General comments and edits. Review: review and make changes to Section 1, 5 and 6. Update figures 1 and 5.

- (BD) David Boulesteix: Wrote, edited, and referenced Sections 2.2 and 5.1.3. Edited/Reviewed 2.1.1, 2.1.2, and 2.4. Creation of Fig. 1 (with Silke Asche) and 2 (analytical part). Reviewed all of Section 2 (content, references, organisation).

- (CEl) Eloi Camprubi: Contributed to general editing and to Sections 2.3 (microfluidics), 2.4.3 (electron microscopy), and 4.1.1 (quantum chemistry). CEl thanks the UT System for a STARs award.

- (CEn) Enrico Sandro Colizzi: Wrote 4.1.1 Replicator models

- (CRA) Alexandre Champagne-Ruel: Contributed Section 4.5, added to Sections 4.4.1, 4.4.2, and 4.3. CRA would like to acknowledge funding from the Natural Sciences and Engineering Research Council of Canada (grant number RGPIN/05278-2018), the Fonds de recherche Nature et Technologies of Québec (grant number 314488), the Fondation J. Armand Bombardier Excellence Scholarship.

- (CSS) Stephanie Colón-Santos: Organised the first draft.

- (HD) Hannah Dromiack: Reviewed Section 4.5.

- (EV) Valentina Erastova: Wrote the core of molecular modelling Section 4.1 (Intro, QM, MD, Figure 3), small contribution to the spectroscopy Sections 2.1 and 2.2 and Section 1.

- (GA) Amanda Garcia: Contributed Significantly to 4.6 Molecular Phylogenetics with particularly significant contributions to ASR

- (GG) Ghjuvan Grimaud: Contributed to Sections 4.3 and 4.4. Contributed Entire section 4.4.3 Whole cell models.

- (HA) Aaron Halpern: Contributed to Section 4.1.

- (HS) Stuart A Harrison: Contributed to Section 2.1.1 UV-Vis Spectroscopy

- (JS) Seán F. Jordan: Reviewed, edited, and contributed to the writing of Section 2. JS acknowledges support from "la Caixa" Foundation (ID 100010434) and from the European Union's Horizon 2020 research and innovation programme under the Marie Skłodowska Curie grant agreement No 847648. The fellowship code is "LCF/BQ/PI21/11830015".

- (JT) Tony Z. Jia: Contributed to 2.4 (Microscopy Techniques), especially 2.4.1 (Light and Fluorescence Microscopy) and 2.4.2 (Confocal Microscopy); contributed to 2.5 (Genomic Sequencing), especially 2.5.1 (Sanger Sequencing) and 2.5.2 (Next Generation Sequencing). Contributed one figure in the Evolution section. JT is supported by Japan Society for the Promotion of Science (JSPS) Grants-in-aid 18K14354 and 21K14746, a Tokyo Institute of Technology Yoshinori Ohsumi Fund for Fundamental Research, the Mizuho Foundation for the Promotion of Sciences and by the Assistant Staffing Program by the Gender Equality Section, Diversity Promotion Office, Tokyo Institute of Technology.

- (KAm) Amit Kahana: contributed to Section 4.1 Molecular Modelling and Simulations, Section 4.4 Complex Systems Modelling, intro of Section 5, Section 5.2 Automation of laboratory experiments.

- (KAr) Artemy Kolchinsky: Wrote Section 4.2.2. Contributed to Section 4.4.1. Helped review/edit Sections 4.1, 4.3, and 4.4. KAr has received funding from the European Union's Horizon 2020 research and innovation programme under the Marie Skłodowska-Curie Grant Agreement No 101068029.

- (MC) Cole Mathis (cole.mathis.ool@gmail.com) Coordinated the writing process, organised the first draft, edited the entire manuscript and handled the submission. He contributed text to the abstract, Section 1, Section 4.3, 4.4, and 4.5. MC would like to thank support from NASA through the Postdoctoral Fellowship Program. The views and conclusions contained in this document are those of the authors and should not be interpreted as representing the official policies, either expressed or implied, of NASA.

- (MGO) Odin Moron-Garcia: Contributed to Section 4.6 Molecular phylogenetics.

- (MO) Omer Markovitch: Contributions on Sections 4, 4.2.2, 4.2.3, 4.3, 4.4, & Abstract; Editing the whole manuscript; General comments, edits and additions.

- **(MR)** Ryo Mizuuchi: Contributed original drafts for Section 5.3, 5.4; editing for Sections 2.5 and 4.4.1.

- **(NJ)** Jingbo Nan: Contributed to Section 2.4.3

- **(OY)** Yuliia Orlova: Contributed Section 4.3.2

- **(PB)** Ben K. D. Pearce: Contributed to Sections 4.2, 4.2.2, 4.2.3. PB is supported by the NSERC Banting Postdoctoral Fellowship.

- **(PK)** Klaus Paschek: Contributed original draft for Section 4.2.1. PK acknowledges the financial support by the Deutsche Forschungsgemeinschaft (**DFG**, German Research Foundation) under Germany's Excellence Strategy EXC 2181/1 - 390900948 (the Heidelberg **STRUCTURES** Excellence Cluster). K.P. is a fellow of the International Max Planck Research School for Astronomy and Cosmic Physics at the University of Heidelberg (**IMPRS-HD**).

- **(PM)** Martina Preiner: Contributed to Section 2.4.3 and overall editing of Section 2.

- **(PS)** Silvana Pinna: Reviewed, edited, and contributed to the writing of Sections 2.1 and 2.2, and contributed to the general editing of Section 2.

- **(PZ)** Zhen Peng: Contributed Section 3 and Section 4.3.3. Network Autocatalysis. Reviewed and made edits to other sections.

- **(RRE)** Eduardo Rodríguez-Román: Contributed to Section 4.6 (Phylogenetics)

- **(SL)** Loraine Schwander: Reviewed, edited, and made contributions to Sections 2.1, 2.2, 2.4, 3, and 4.6. Contributed to section 5.2.

- **(SS)** Siddhant Sharma: Contributed to 2.6.2 Raman Spectroscopy, 3.1 Physical and Chemical Data, 3.2 Biochemical and Biological Databases, and 4.3.2 Automated Reaction Network Generation. S.S. would like to acknowledge the SETI Forward 2020 from the SETI Institute.

- **(VA)** Andrey Vieira: Wrote the first draft, made some figures, organised references, and edited the manuscript.

- **(XJ)** Joana C. Xavier: Helped organise the first draft. Editing the whole manuscript. Contributed to Sections 1 2, 3 and Section 4.3.

equilibrium furanose selection in the ribose isomerisation network. Nat Commun. 2021;12: 2749.